\documentclass[a4paper,11pt]{article}
\pdfoutput=1 
\usepackage{graphicx,amsmath,amsfonts,amssymb,dcolumn,amsthm,euscript}
\makeatletter
\@addtoreset{equation}{section}

\makeatother

\newcommand\bg{\bar g}
\newcommand\bR{\bar R}
\newcommand\bnabla{\bar\nabla}
\newcommand\bgamma{\bar\gamma}
\newcommand\bea{\begin{eqnarray}}
\newcommand\eea{\end{eqnarray}}
\newcommand\be{\begin{equation}}
\newcommand\ee{\end{equation}}
\newcommand\tr{\mathrm{tr}}
\newcommand\Tr{\mathrm{Tr}}
\newcommand{\vs}[1]{\vspace{#1 mm}}

\begin{document}

\begin{titlepage}

\renewcommand{\thefootnote}{\fnsymbol{footnote}}
\begin{flushright}
KU-TP 066 \\
\today
\end{flushright}

\vs{10}
\begin{center}
{\Large\bf Gauges and functional measures in quantum gravity
 I:\\ Einstein theory}
\vs{15}

{\large
N. Ohta,$^{a,}$\footnote{e-mail address: ohtan@phys.kindai.ac.jp}
R. Percacci$^{b,c,}$\footnote{e-mail address: percacci@sissa.it}
and A. D. Pereira$^{b,d,e,}$\footnote{e-mail address: aduarte@if.uff.br}
} \\
\vs{10}
$^a${\em Department of Physics, Kindai University,
Higashi-Osaka, Osaka 577-8502, Japan}

$^b${\em International School for Advanced Studies, via Bonomea 265, 34136 Trieste, Italy}

$^c${\em INFN, Sezione di Trieste, Italy}

$^d${\em Universidade Federal Fluminense, Instituto de F\'{\i}sica, Campus da Praia Vermelha, \\
Avenida General Milton Tavares de Souza s/n, 24210-346, Niter\'oi, RJ, Brasil.}

$^e${\em Max Planck Institute for Gravitational Physics (Albert Einstein Institute)\\
Am M\"uhlenberg 1, Potsdam, 14476 Germany}

\vs{15}
{\bf Abstract}
\end{center}

We perform a general computation of the off-shell one-loop divergences in Einstein gravity,
 in a two-parameter family of path integral measures, corresponding to different ways of parametrizing
the graviton field, and a two-parameter family of gauges.
Trying to reduce the gauge- and measure-dependence selects
certain classes of measures and gauges respectively.
There is a choice of two parameters
(corresponding to the exponential parametrization
and the partial gauge condition that the quantum field be traceless)
that automatically eliminates the dependence 
on the remaining two parameters and on the cosmological constant.
We observe that the divergences are invariant under a $\mathbf{Z}_2$ ``duality'' transformation that 
(in a particularly important special case)
involves the replacement of the densitized metric
by a densitized inverse metric as the fundamental quantum variable.
This singles out a formulation of unimodular gravity
as the unique ``self-dual'' theory in this class.

\end{titlepage}
\newpage
\setcounter{page}{2}
\renewcommand{\thefootnote}{\arabic{footnote}}
\setcounter{footnote}{0}

\section{Introduction}

If one properly takes into account the Jacobian of the transformation,
then a change of parametrization of the degrees of freedom of a system
should leave physical observables unchanged, on-shell.
In quantum field theory, this is known as the equivalence theorem
\cite{Lam:1973qa,Tyutin:2000ht,Ferrari:2002kz}.
On the other hand, if one does not take the Jacobian into account,
functional integrals written in terms of different variables
correspond to different choices of the functional measure.
In principle these define different quantum theories and may well yield
different results even for physical observables.
Insofar as the functional integral is a purely formal expression,
such general statements have to be verified after introducing
suitable regularizations and renormalizations.

In quantum gravity there are many possible choices of variables,
and of functional measure.
Even restricting our attention to formulations of the theory in terms
of a metric alone, one can take as fundamental quantum field the
densitized metric
\be
\gamma_{\mu\nu}=g_{\mu\nu}\left(\sqrt{\det g_{\mu\nu}}\right)^w ,
\label{denslin}
\ee
or the densitized inverse metric
\be
\gamma^{\mu\nu}
=g^{\mu\nu}\left(\sqrt{\det g_{\mu\nu}}\right)^w\ ,
\label{densinv}
\ee
where $w$ is known as the weight (note that here $\gamma_{\mu\nu}$ and
$\gamma^{\mu\nu}$ have the same weight and are not
the inverse of each other).
Writing the functional integral in terms of $\gamma_{\mu\nu}$
instead of $g_{\mu\nu}$ amounts to an ultralocal change in the
functional measure.
Several authors have suggested specific choices of $w$
or even different, non-covariant measures
\cite{Misner:1957wq,Leutwyler:1964wn,DeWitt:1965jb,Faddeev:1973zbs,Fradkin:1974df,
Fujikawa:1983im,Fujikawa:1984qk,Anselmi:1991wb}.
Here we will not commit to any such choice but treat $w$
as a free parameter.

The definition of the functional integral of quantum gravity
as an integral over metrics is still rather formal.
In quantum gravity it is almost inevitable to use the background field method
where the true quantum variable is not the (densitized) metric
but rather its deviation from some classical value.
This gives rise to another ambiguity: the densitized metric
can be written in the form
$\gamma=f(\bar\gamma,\hat h)$,
where the function $f$ has the property
$f(\bar\gamma,0)=\bar\gamma$.
The most common procedure is to expand the (densitized) metric linearly
\be
\label{linexp}
\gamma_{\mu\nu}=\bar\gamma_{\mu\nu}+\hat h_{\mu\nu}
\qquad \mathrm{or}\qquad
\gamma^{\mu\nu}=\bar\gamma^{\mu\nu}+\hat h^{\mu\nu}\ ,
\ee
but in the literature the exponential form
\be
\label{expexp}
\gamma_{\mu\nu}=\bar\gamma_{\mu\rho}(e^{\hat h})^\rho{}_\nu
\qquad \mathrm{or}\qquad
\gamma^{\mu\nu}=(e^{-\hat h})^\mu{}_\rho\bgamma^{\rho\nu}\ ,
\ee
has also been used
\cite{Kawai:1989yh,pv1,nink}.
We will see that, at least at the one-loop level, all these discrete
choices can be subsumed in a continuous parameter $\omega$.

In gravity, as in any gauge theory,
there are further ambiguities due to the need
of introducing a gauge fixing procedure.
On-shell quantities will generally be independent of
the choice of gauge, but sometimes one is interested
in off-shell quantities.
As an example we may cite the calculation of one-loop divergences
in quantum gravity, where it can be shown that there exists
a gauge choice for which the coefficient of the
logarithmic divergences is zero \cite{kallosh}.
Off-shell calculations of beta functions are common in the asymptotic safety approach to quantum gravity \cite{reviews},
and in such cases one may be interested in minimizing
their dependence on the gauge choice.

In this paper we consider one-loop corrections in quantum general relativity (GR),
and in particular the coefficient of the leading divergences
(in four dimensions: quartic, quadratic and logarithmic divergences).
We will study the dependence of these coefficients
on two gauge parameters and a two-parameter family of
functional measures.
Similar calculations in a less general setting have been
performed in \cite{Kalmykov:1995fd}.

The plan of the paper is as follows. In section 2 we define the parametrization of the metric
In section 3 we describe the choice of gauge and the calculation of
the one-loop effective action.
Section 4 contains the discussion of the results for the one-loop divergences.
In section 5 we discuss the duality between measures
and section 6 contains a brief discussion.

We plan to extend these result to higher derivative gravity
in  forthcoming publication.

\section{Parametrization of the quantum fluctuations}

Our starting point is the gravitational action $S(g)$,
written in terms
of a metric $g_{\mu\nu}$ in $d$ Euclidean dimensions.
We will assume that the fundamental dynamical variable is not
the metric itself but rather a tensor density $\gamma_{\mu\nu}$
or $\gamma^{\mu\nu}$ of weight $w$.
The metric and its inverse are defined by
\begin{equation}
g_{\mu\nu}=\gamma_{\mu\nu}\left(\det(\gamma_{\mu\nu})\right)^{m}
\ ;\qquad
g^{\mu\nu}=\gamma^{\mu\nu}\left(\det(\gamma_{\mu\nu})\right)^{-m}
\label{fc2}
\end{equation}
where $\gamma^{\mu\alpha}\gamma_{\alpha\nu}
=g^{\mu\alpha}g_{\alpha\nu}=\delta^{\mu}_{\nu}$.
This implies a relation between the determinants of $g$
and $\gamma$, namely
\begin{equation}
\det g=\left(\det\gamma\right)^{1+dm}\,.
\label{fc3}
\end{equation}
For $m\not=-1/d$, the relations (\ref{fc2}) can be inverted:
\be
\gamma_{\mu\nu}=g_{\mu\nu}(\det(g_{\mu\nu}))^{-\frac{m}{1+dm}}
\ ,\qquad
\gamma^{\mu\nu}=g^{\mu\nu}(\det(g_{\mu\nu}))^{\frac{m}{1+dm}}\ .
\ee
Comparing with (\ref{denslin}) and (\ref{densinv})
we find that $m$ is related to the weight by
\be
\frac{w}{2}=-\frac{m}{1+dm}\qquad
\mathrm{or}\qquad
\frac{w}{2}=\frac{m}{1+dm}\ ,
\label{relation}
\ee
respectively.
Conversely, $m=-\frac{w/2}{1+dw/2}$ for (\ref{denslin})
and $m=\frac{w/2}{1-dw/2}$ for (\ref{densinv}).
We observe that the relation between $m$ and $w/2$ is an involution.
We choose to treat $m$ as an independent free parameter.
All dependence on $m$ can be translated into a dependence on $w$
if needed, using the preceding formulas.

For $m=-1/d$, the transformation (\ref{fc2}) is singular and from eq.(\ref{fc3}), we see that it implies that the determinant of $g_{\mu\nu}$ is one.
We will refer to this choice as unimodular gravity.
For this reason, quantum corrections in
this specific case should be analyzed separately.

For the calculation of one-loop effects one needs the
expansion of the action around a background field.
This will now depend on the parametrization of the metric.
For both cases (\ref{linexp}) and (\ref{expexp}),
if we momentarily use $\bar \gamma_{\mu\nu}$
and its inverse $\bar \gamma^{\mu\nu}$
to raise and lower indices,
the field $\hat h^\mu{}_\nu$ is a genuine tensor and
the fields $\hat h_{\mu\nu}=\bar\gamma_{\mu\rho}\hat h^\rho{}_\nu$
and
$\hat h^{\mu\nu}=\hat h^\mu{}_\rho\bar\gamma^{\rho\nu}$
are densities of the same weight as $\gamma_{\mu\nu}$
and $\gamma^{\mu\nu}$.
It is preferable to work with a quantum fluctuation
that is a true tensor, so we define
\be
h_{\mu\nu}=(\det\bar\gamma)^m\hat h_{\mu\nu}\ .
\ee
Now that we have a genuine tensor,
we can avoid having to deal with explicit powers of determinants
by using the background metric
\be
\bg_{\mu\nu}=\bgamma_{\mu\nu}(\det\bgamma)^m\ ;\qquad
\bg^{\mu\nu}=\bgamma^{\mu\nu}(\det\bgamma)^{-m},
\ee
to raise and lower indices.
For example
\be
\label{redefh}
h^{\mu\nu}=\bg^{\mu\rho}\bg^{\nu\sigma}h_{\rho\sigma}
=(\det\bar\gamma)^{-m}\bgamma^{\mu\rho}\bgamma^{\nu\sigma}\hat h_{\rho\sigma}
=(\det\bar\gamma)^{-m}\hat h^{\mu\nu}\ .
\ee

We can now write
\be
\label{gexp}
g_{\mu\nu}=\bg_{\mu\nu}+\delta g_{\mu\nu}\ ,
\ee
and the fluctuation can be expanded:
\be
\label{gammaexp}
\delta g_{\mu\nu}=
\delta g^{(1)}_{\mu\nu}
+\delta g^{(2)}_{\mu\nu}
+\delta g^{(3)}_{\mu\nu}+\ldots \ ,
\ee
where $\delta g^{(n)}_{\mu\nu}$ contains $n$
powers of $h_{\mu\nu}$.

Let us begin from the case where the quantum field is the
densitized covariant metric $\gamma_{\mu\nu}$ and we use
the linear background field expansion (\ref{linexp}).
For the expansion of the determinant of $\gamma_{\mu\nu}$
one writes
\be
\det\gamma_{\mu\nu}=\det(\bgamma_{\mu\rho}(\delta^\rho_\nu+\hat h^\rho{}_\nu))
=\det(\bgamma)\det(\mathbf{1}+\hat h)\ ,
\ee
and then expands
\be
\det(\mathbf{1}+\hat h)
=\exp\tr\log(\mathbf{1}+\hat h)
=\exp\sum_{n=1}^\infty \frac{(-1)^{n+1}}{n}\tr\hat h^n
=e^{h_1}
e^{-\frac{1}{2}h_2}
e^{\frac{1}{3}h_3}
e^{-\frac{1}{4}h_4}\ldots \ ,
\ee
where $h_n=\tr\hat h^n$, {\it e.g.}
$h_1=\hat h^\mu{}_\mu=h^\mu{}_\mu\equiv h$,
$h_2=h^\alpha{}_\beta h^\beta{}_\alpha$,
$h_3=h^\alpha{}_\beta h^\beta{}_\gamma h^\gamma{}_\alpha$ etc..
Here and everywhere in the following indices will be raised and
lowered with the background metric $\bg^{\mu\nu}$, $\bg_{\mu\nu}$.
This leads to the expansion
\bea
\label{lindir}
g_{\mu\nu}&=&\bar g_{\mu\nu}
+h_{\mu\nu}
+m\bar g_{\mu\nu}h
+m h h_{\mu\nu}
+\frac{1}{2}\bar g_{\mu\nu}(-mh_2+m^2h^2)
\nonumber\\
&+&\frac{1}{2}h_{\mu\nu}(-mh_2+m^2 h^2)
+\bg_{\mu\nu}\left(\frac{m}{3}h_3
-\frac{m^2}{2}h h_2
+\frac{m^3}{6}h^3
\right)
\\
&+&h_{\mu\nu}\left(\frac{m}{3}h_3
-\frac{m^2}{2}h h_2
+\frac{m^3}{6}h^3\right)
+\bg_{\mu\nu}\left(
-\frac{m}{4}h_4
+\frac{m^2}{3}h h_3
+\frac{m^2}{8}h_2^2
-\frac{m^3}{4}h^2 h_2
+\frac{m^4}{24}h^4
\right)+\ldots\ .
\nonumber
\eea

Instead, if the quantum field is the densitized inverse metric
$\gamma^{\mu\nu}$ of weight $2m/(1+dm)$, the linear expansion
\be
\gamma^{\mu\nu}=\bar\gamma^{\mu\nu}-\hat h^{\mu\nu} \ ,
\label{invlinsplit1}
\ee
(notice the minus sign),
followed by the redefinition (\ref{redefh}), leads to
%
%
\bea
\label{lininv}
g_{\mu\nu}&=&\bar g_{\mu\nu}
+h_{\mu\nu}
+m\bar g_{\mu\nu}h
+h_{\mu\rho}h^\rho{}_\nu
+m h h_{\mu\nu}
+\frac{1}{2}\bar g_{\mu\nu}(mh_2+m^2h^2)
\nonumber\\
&+&h_{\mu\rho}h^\rho{}_\sigma h^\sigma{}_\nu
+m h h_{\mu\rho}h^\rho{}_\nu
+\frac{1}{2}h_{\mu\nu}(mh_2+m^2h^2)
+\bg_{\mu\nu}\left(\frac{m}{3}h_3
+\frac{m^2}{2}h h_2
+\frac{m^3}{6}h^3\right)
\\
&+&h_{\mu\rho}h^\rho{}_\sigma h^\sigma{}_\lambda h^\lambda{}_\nu
+mh h_{\mu\rho}h^\rho{}_\sigma h^\sigma{}_\nu
+\frac{1}{2}h_{\mu\rho}h^\rho{}_\nu(mh_2+m^2h^2)
\nonumber\\
&+&h_{\mu\nu}\left(\frac{m}{3}h_3
+\frac{m^2}{2}h h_2
+\frac{m^3}{6}h^3\right)
+\bg_{\mu\nu}\left(\frac{m}{4}h_4
+\frac{m^2}{3}h h_3
+\frac{m^2}{8}h_2^2
+\frac{m^3}{4}h^2 h_2
+\frac{m^4}{24}h^4\right)+\ldots\ .
\nonumber
\eea

Now consider the exponential expansion of the densitized metric
as defined in (\ref{expexp}).
In this case the expansion of the determinant only produces
terms proportional to the single trace $h$:
$$
\det(\gamma_{\mu\nu})
=\det(\bgamma_{\mu\nu})\det e^{\hat h}
=\det(\bgamma_{\mu\nu})e^{\tr\hat h}
=\det(\bgamma_{\mu\nu})\left(1+h+\frac{1}{2}h^2+\frac{1}{3!}h^3+\ldots\right)\ .
$$
The exponential expansion (\ref{expexp}),
followed by the redefinition (\ref{redefh}), leads to
\bea
\label{expdir}
g_{\mu\nu}&=&\bar g_{\mu\nu}
+h_{\mu\nu}
+m\bar g_{\mu\nu}h
+\frac{1}{2!}h_{\mu\rho}h^\rho{}_\nu
+m h h_{\mu\nu}
+\frac{m^2}{2}\bar g_{\mu\nu}h^2
\nonumber\\
&+&\frac{1}{3!}h_{\mu\rho}h^\rho{}_\sigma h^\sigma{}_\nu
+\frac{m}{2} h h_{\mu\rho}h^\rho{}_\nu
+\frac{m^2}{2}h^2 h_{\mu\nu}
+\frac{m^3}{3!}h^3\bg_{\mu\nu}
\\
&+&\frac{1}{4!}h_{\mu\rho}h^\rho{}_\sigma h^\sigma{}_\lambda h^\lambda{}_\nu
+\frac{m}{3!}h h_{\mu\rho}h^\rho{}_\sigma h^\sigma{}_\nu
+\frac{m^2}{4}h^2 h_{\mu\rho}h^\rho{}_\nu
+\frac{m^3}{3!}h^3 h_{\mu\nu}
+\frac{m^4}{4!}h^4\bg_{\mu\nu}+\ldots\ .
\nonumber
\eea
Finally, the exponential expansion of the inverse metric
\be
\gamma^{\mu\nu}=(e^{-\hat h})^\mu{}_\rho\bgamma^{\rho\nu}\ ,
\ee
leads again to the same formula (\ref{expdir}).

For the one-loop evaluation of the effective action
we only need the expansions up to second order in the fluctuation,
which are contained in the first lines of
(\ref{lindir},\ref{lininv},\ref{expdir}).
We observe that they are special cases of a
two-parameter family of expansions of the form (\ref{gammaexp}),
with
\bea
\delta g^{(1)}_{\mu\nu}&=&h_{\mu\nu}+m\bg_{\mu\nu}h \ ,
\nonumber\\
\delta g^{(2)}_{\mu\nu}&=&
\omega h_{\mu\rho}h^\rho{}_\nu
+m h h_{\mu\nu}
+m\left(\omega-\frac{1}{2}\right)\bg_{\mu\nu}h^{\alpha\beta}h_{\alpha\beta}
+\frac{1}{2}m^2\bg_{\mu\nu}h^2\ .
\label{deltag}
\eea
Here the choice $\omega=0$ corresponds to the
linear expansion of metric,
$\omega=1/2$ corresponds to the exponential expansion
and $\omega=1$ corresponds to the
linear expansion of the inverse metric, as in eq.(\ref{invlinsplit1}).
(As a matter of fact, one observes that to this order the exponential
expansion is just the mean of the other two.)

\section{One-loop quantum GR}

We are going to calculate the formal path integrals
\be
\int \left[\EuScript{D}\gamma_{\mu\nu}\right]\,\mathrm{e}^{-S(g(\gamma))}
\qquad \mathrm{and} \qquad
\int \left[\EuScript{D}\gamma^{\mu\nu}\right]\,\mathrm{e}^{-S(g(\gamma))} \ ,
\ee
where the action $S(g)$ is kept the same, but is rewritten in terms
of the quantum fields $\gamma_{\mu\nu}$ or $\gamma^{\mu\nu}$
using equation (\ref{fc2}),
and $\left[\EuScript{D}\gamma_{\mu\nu}\right]$, $\left[\EuScript{D}\gamma^{\mu\nu}\right]$ denote
the usual translation-invariant
functional measures for $\gamma_{\mu\nu}$ or $\gamma^{\mu\nu}$.
We are thus going to repeat the classic calculation of
\cite{'tHooft:1974bx}, but in a more general context:
in any dimension, in a two-parameter family of gauges
specified below, in the two-parameter family of measures
specified above, and also keeping track of the
leading (power) divergences.

\subsection{Expansion of the action}

We now concentrate on the Hilbert action
\bea
S(g(\gamma))&=&Z_N\int d^dx\sqrt{g}
(2\Lambda -g^{\mu\nu}R_{\mu\nu}(g))
\nonumber \\
&=&Z_N\int d^dx~
(\det\gamma)^{\frac{1+dm}{2}}
\left(2\Lambda-
(\det\gamma)^{-m}\gamma^{\mu\nu}R_{\mu\nu}(g(\gamma))\right)\,.
\label{fc1}
\eea
where $Z_N=1/(16\pi G)$ and
$\Lambda$ and $G$ denote the cosmological and Newton
constants, respectively.

The expansion of the action to second order in the quantum
fluctuation $h_{\mu\nu}$ can be obtained as follows.
One begins with the standard expansion of the Hilbert action,
regarded as a function of the metric $g_{\mu\nu}$,
to second order in $\delta g_{\mu\nu}$:
\be
S(g)=S(\bg)
+\int d^dx\,\sqrt{\bg}\, E^{\mu\nu}\delta g_{\mu\nu}
+\frac{1}{2}\int d^dx\,\sqrt{\bg}\,\delta g_{\mu\nu}
H^{\mu\nu\rho\sigma}\delta g_{\rho\sigma}
+\ldots \ .
\ee
Then one replaces $\delta g_{\mu\nu}$
by its expansion (\ref{gammaexp}) to second order in $h_{\mu\nu}$
to obtain
\bea
S(g(\gamma))&=&S(\bg)
+\int d^dx\,\sqrt{\bg}\,
E^{\mu\nu}(\delta g^{(1)}_{\mu\nu}(m)+\delta g^{(2)}_{\mu\nu}(m,\omega))
\nonumber\\
&&+\frac{1}{2}\int d^dx\,\sqrt{\bg}\, \delta g^{(1)}_{\mu\nu}(m)
H^{\mu\nu\rho\sigma}
\delta g^{(1)}_{\rho\sigma}(m)
+\ldots
\nonumber
\\
&=&S(\bg)
+\int d^dx\,\sqrt{\bg}\,
E^{\prime\mu\nu}(m)h_{\mu\nu}
\nonumber\\
&&+\frac{1}{2}\int d^dx\,\sqrt{\bg}\, h_{\mu\nu}
H^{\prime\mu\nu\rho\sigma}(m,\omega)
h_{\rho\sigma}
+\ldots \ .
\nonumber
\eea
The modified Hessian $H^{\prime\mu\nu\rho\sigma}$
contains terms coming from the equation of motion.

Expanding around a maximally symmetric background,
with curvature tensor
\begin{equation}
\bR_{\mu\nu\alpha\beta}=
\frac{\bR}{d(d-1)}
(\bg_{\mu\alpha}\bg_{\nu\beta}-\bg_{\mu\beta}\bg_{\nu\alpha})
\,,
\label{fc13}
\end{equation}
this procedure leads to the following quadratic action
\begin{eqnarray}
{S}^{(2)}&=&\frac{Z_N}{2}
\int d^dx\,\sqrt{\bg}\Bigg\{
\frac{1}{2}h_{\mu\nu}(-\bar\nabla^2)h^{\mu\nu}
+h_{\mu\nu}\bar\nabla^\mu\bar\nabla^\rho h_\rho{}^\nu
-\left(1+(d-2)m\right)h\bar{\nabla}^{\mu}\bar{\nabla}^{\nu}h_{\mu\nu}
\nonumber\\
&&+\frac{1}{2}\left(1+2(d-2)m+(d-1)(d-2)m^2)\right)h\bar{\nabla}^{2}h
\nonumber\\
&&
+\left[\frac{\bR}{d(d-1)}
-(1+dm)(1-2\omega)\left(\Lambda-\frac{d-2}{2d}\bR\right)\right]
h^{\mu\nu}h_{\mu\nu}
\nonumber\\
&&
+\left[
\frac{d-3+m(d-1)(d-2)(1+dm)}{2d(d-1)}\bR
+\frac{(1+dm)^2}{2}\left(\Lambda-\frac{d-2}{2d}\bR\right)
\right]
h^2
\Bigg\}\,.
\label{feyg2}
\end{eqnarray}
The bars on the covariant derivatives means that they are calculated
from the background metric $\bg_{\mu\nu}$. 
Notice that $\omega$ only appears in the third line.

We note that the most general form of the quadratic term
in (\ref{deltag}) would be
\be
\delta g^{(2)}_{\mu\nu}=\frac{1}{2}\left(
\tau_1 h_{\mu\rho}h^\rho{}_\nu
+\tau_2 h h_{\mu\nu}
+\tau_3\bg_{\mu\nu}h^{\alpha\beta}h_{\alpha\beta}
+\tau_4\bg_{\mu\nu} h^2
\right) .
\ee
As already observed in \cite{Gies:2015tca}, these parameters appear in the
expansion of the Hilbert action only through the combinations
$T_1=\frac{1}{4}\tau_1+\tau_3$ and $T_2=\frac{1}{4}\tau_2+\tau_4$.
These are related to our parameters $m$ and $\omega$ by
\be
T_1=\frac{1}{2}\omega(1+4m)-m\ ;\qquad
T_2=\frac{1}{2}m(1+2m)\ .
\ee

\subsection{Gauge fixing and ghosts}

We consider a general linear background gauge-fixing condition
\begin{equation}
F_{\mu}=\bar{\nabla}_{\alpha}{h^{\alpha}}_{\mu}-\frac{\bar b+1}{d}\bar{\nabla}_{\mu}h\,,
\label{gaugecondition}
\end{equation}
depending on a parameter $-\infty<\bar b<\infty$,
where $h_{\mu\nu}$ is the tensorial quantum field defined above.
The gauge-fixing term in the action is
\begin{equation}
S_{GF}=\frac{Z_N}{2a}\int d^dx\,\sqrt{\bg}\,\bg^{\mu\nu}F_{\mu}F_{\nu}\,,
\label{gfa1}
\end{equation}
where $a$ is a gauge parameter.
The usual harmonic (de Donder) gauge condition
corresponds to $\bar b=\frac{d}{2}-1$.
The gauge parameter $a$ is assumed to be positive or zero.
The choice $a=1$ (Feynman gauge) is often used because
it simplifies calculations greatly.
While on-shell the choice of $a$ should be completely immaterial,
we note that the effect of the unphysical (gauge) degrees of freedom
is more suppressed the smaller $a$ is, so that in some sense
the Landau gauge $a=0$, which amounts to
imposing the gauge condition strongly,
is expected to give the most reliable results.
When the gauge parameters are allowed to run with scale,
$a=0$ is expected to be a fixed point
\cite{Ellwanger:1995qf}.

For reasons that will become apparent later,
it will be convenient to redefine the gauge parameter
\be
\bar b=b(1+dm)\ .
\label{betaredef}
\ee
After an integration by parts, the gauge fixing term can be written as
\begin{equation}
S_{GF}=-\frac{Z_N}{2a}\int d^dx\,\sqrt{\bg}
\left[h_{\mu\nu}\bar{\nabla}^{\nu}\bar{\nabla}^{\alpha}{h_{\alpha}}^{\mu}
-2\frac{1+b(1+dm)}{d}h\bar{\nabla}^{\mu}\bar{\nabla}^{\nu}h_{\mu\nu}
+\left(\frac{1+b(1+dm)}{d}\right)^2h\bar{\nabla}^{2}h\right]\,.
\label{gfa2}
\end{equation}

Some care is required in the derivation of the ghost action.
Although we have found it convenient to rewrite the expansion of the
action in terms of the tensorial variable $h_{\mu\nu}$,
in the Faddeev-Popov procedure one has to recall that the
quantum field is the tensor density $\gamma_{\mu\nu}$,
and it is the infinitesimal gauge transformation of this
quantity that enters in the definition of the Faddeev-Popov determinant.
The infinitesimal gauge variation of $\gamma_{\mu\nu}$ is
\begin{equation}
\mathcal{L}_\epsilon\gamma_{\mu\nu}=
\gamma_{\mu\rho}\nabla_{\nu}\epsilon^\rho
+\gamma_{\nu\rho}\nabla_{\mu}\epsilon^\rho
-\frac{2m}{1+dm}\gamma_{\mu\nu}\nabla_\lambda\epsilon^\lambda
\ .
\label{gh01}
\end{equation}
When converted into a tensor, this gives an infinitesimal variation
\begin{equation}
h^\epsilon_{\mu\nu}=\nabla_\mu\epsilon_\nu+\nabla_\nu\epsilon_\mu
-\frac{2m}{1+dm}g_{\mu\nu}\nabla_\lambda\epsilon^\lambda\ ,
\end{equation}
where indices have been lowered with the metric $g_{\mu\nu}$.
The Faddeev-Popov ghost is obtained by inserting this
gauge variation in the gauge condition.
A short algebra leads to the ghost action
\begin{eqnarray}
S_{\mathrm{gh}}&=&-\int d^dx\,\sqrt{\bg}\,\bar{C}^{\mu}\frac{\partial F_{\mu}}{\partial \hat{h}_{\alpha\beta}}\mathcal{L}_{C}\gamma_{\alpha\beta}
\nonumber\\
&=&-\int d^dx\,\sqrt{\bg}~\bar{C}^{\mu}\left[\delta^{\nu}_{\mu}\bar{\nabla}^{2}+\left(1-2\frac{1+b}{d}\right)\bar{\nabla}_{\mu}\bar{\nabla}^{\nu}+\frac{\bar{R}}{d}\delta^{\nu}_{\mu}\right]C_{\nu}\,,
\label{gha1}
\label{gh0}
\end{eqnarray}
Note that here $b$ appears without the factor $1+dm$ that
is ubiquitous elsewhere.

\subsection{York Decomposition}

Following \cite{Fradkin:1983mq},
the Hessian can be nearly diagonalized by using the York
decomposition of the fluctuation field:
\begin{equation}
h_{\mu\nu}=h^{\mathrm{TT}}_{\mu\nu}+\bar{\nabla}_{\mu}\xi_{\nu}+\bar{\nabla}_{\nu}\xi_{\mu}+\bar{\nabla}_{\mu}\bar{\nabla}_{\nu}\sigma-\frac{1}{d}\bar{g}_{\mu\nu}\bar{\nabla}^{2}\sigma+\frac{1}{d}\bar{g}_{\mu\nu}h\,,
\label{ap1}
\end{equation}
where
\begin{equation}
\bar{\nabla}^{\mu}h^{\mathrm{TT}}_{\mu\nu}=0\,,\,\,\,\, \bar{g}^{\mu\nu}h^{\mathrm{TT}}_{\mu\nu}=0\,,\,\,\,\, \bar{\nabla}^{\mu}\xi_{\mu}=0\,,\,\,\,\, h=\bar{g}^{\mu\nu}h_{\mu\nu}\,.
\label{ap2}
\end{equation}

It is convenient to redefine the fields $\xi_{\mu}$ and $\sigma$
so that they have the same dimension as $h_{\mu\nu}$:
\be
\hat{\xi}_{\mu}=\sqrt{-\bar{\nabla}^{2}-\frac{\bar{R}}{d}}\xi_{\mu}
\ \ ;\qquad
\hat{\sigma}=\sqrt{-\bar{\nabla}^{2}}\sqrt{-\bar{\nabla}^{2}-\frac{\bar{R}}{d-1}}\sigma\,.
\label{ap6}
\ee

The York decomposition leads to a non-trivial Jacobian while redefinitions (\ref{ap6}) produce another Jacobian which exactly cancels the previous one.

After the York decomposition and field redefinition
\begin{eqnarray}
S+S_{GF}\!\!&=&\!\!\frac{Z_N}{2}\!\int d^dx\,\sqrt{\bg}
\Bigg\{
\frac{1}{2}h^{\mathrm{TT}}_{\mu\nu}
\left[-\bnabla^2
+\frac{2\bR}{d(d-1)}
-2(1+dm)(1-2\omega)\left(\Lambda-\frac{d-2}{2d}\bR\right)\right]h^{\mathrm{TT}\mu\nu}
\nonumber\\
&+&
\frac{1}{a}\hat{\xi}_{\mu}
\left[-\bnabla^2-\frac{\bR}{d}
-2a(1+dm)(1-2\omega)\left(\Lambda-\frac{d-2}{2d}\bR\right)
\right]\hat{\xi}^{\mu}
\nonumber\\
&-&
\frac{d-1}{2d}\hat{\sigma}\bigg[
\frac{a(d-2)-2(d-1)}{da}(-\bnabla^2)
+\frac{2\bR}{da}
+2(1+dm)(1-2\omega)
\left(\Lambda-\frac{d-2}{2d}\bar{R}\right)
\bigg]\hat{\sigma}
\nonumber\\
&-&
\frac{(d-1)(1+dm)\big((d-2)a-2b\big)}{d^2a}
\hat\sigma\sqrt{-\bar{\nabla}^{2}}\sqrt{-\bar{\nabla}^{2}
-\frac{\bar{R}}{d-1}}h
\nonumber\\
&-&
h\frac{(1+dm)^2}{2d^2a}\bigg[
\big((d-1)(d-2)a-2b^2\big)(-\bnabla^2)
-(d-2)a\bR
\nonumber\\
&-&da\left(d-2\frac{1-2\omega}{1+dm}\right)
\left(\Lambda-\frac{d-2}{2d}\bR\right)
\bigg]h
\Bigg\}\,.
\label{ehgf1}
\end{eqnarray}

The only residual non-diagonal terms are in the $\sigma$-$h$ sector.

Similarly, when the ghosts are decomposed
in their longitudinal and transverse parts
\be
C_\mu=C^{\mathrm{T}}_{\mu}+\bar{\nabla}_{\mu}\frac{1}{\sqrt{-\bar{\nabla}^{2}}}C'^{L}
\ \ ;\qquad
\bar{C}^{\mu}=\bar{C}^{\mathrm{T}\mu}+\bar{\nabla}^{\mu}\frac{1}{\sqrt{-\bar{\nabla}^{2}}}\bar{C}'^{L},
\label{ghs2}
\ee
the ghost action becomes
\begin{equation}
S_{\mathrm{gh}}=-\int d^dx\,\sqrt{\bg}\left[\bar{C}^{\mathrm{T}\mu}\left(\bar{\nabla}^{2}+\frac{\bar{R}}{d}\right)C^{\mathrm{T}}_{\mu}+2\frac{d-1-b}{d}\bar{C}'^{L}\left(\bar{\nabla}^{2}+\frac{\bar{R}}{d-1-b}\right)C'^{L}\right]\,.
\label{ghfr1}
\end{equation}

\section{One-loop divergences}

The one-loop effective action contains a divergent part
\be
\Gamma_k=\int d^dx\,\sqrt{\bg}\left[
\frac{A_1}{16\pi d} k^d
+\frac{B_1}{16\pi(d-2)} k^{d-2}\bar R
+\frac{C_1}{d-4} k^{d-4}\bar R^2+\ldots\right]\,,
\nonumber
\label{gammaabc}
\ee
where $k$ stands for a cutoff and we introduced
a reference mass scale $\mu$.
In $d=4$, the last term is replaced by $C_1\log(k/\mu)\bR^2$.
In general one would have separate
Riemann squared, Ricci squared and $R^2$ terms,
but here we use the curvature conditions (\ref{fc13})
and reduce them all to a single term proportional to $\bR^2$.

The coefficients
$A_1$, $B_1$, $C_1$ depend on
$d$, $m$, $\omega$, $a$, $b$ and $\tilde\Lambda=k^{-2}\Lambda$.
These functions are too complicated to be reported in generality.
We describe here the algorithm that is used to derive them,
so that the readers can easily reproduce on a computer.
Instead of $\Gamma_k$ we shall evaluate the derivative
\cite{Reuter,Dou}:
\be
\dot\Gamma_k=
\int d^dx\,\sqrt{\bg}\left[
\frac{A_1}{16\pi} k^d
+\frac{B_1}{16\pi} k^{d-2}\bar R
+C_1 k^{d-4}\bar R^2+\ldots\right],
\label{gammadotabc}
\ee
where the dot stands for $k\frac{d}{dk}$.
The one-loop effective action $\Gamma_k$ with cutoff $k$ is given by
\be
\Gamma_k=\frac{1}{2}\Tr\log\left(\frac{\Delta_k^{(2)}}{\mu^2}\right)
+\frac{1}{2}\Tr\log\left(\frac{\Delta_k^{(1)}}{\mu^2}\right)
+\frac{1}{2}\Tr\log\left(\frac{\Delta_k^{(0)}}{\mu^2}\right)
-\Tr\log\left(\frac{\Delta_{gh,k}^{(1)}}{\mu^2}\right)
-\Tr\log\left(\frac{\Delta_{gh,k}^{(0)}}{\mu^2}\right)\ .
\ee
where each $\Delta_k$ is one of the kinetic operators
that appear in (\ref{ehgf1}),
in which the Bochner Laplacian $-\bnabla^2$ has been replaced by
$P_k(-\bnabla^2)=-\bnabla^2+R_k(-\bnabla^2)$.
The kernel $R_k(-\bnabla^2)$ is to some extent arbitrary,
but its effect must be to suppress the contribution of the modes
with eigenvalues below $k^2$. Thus, it must go to zero
sufficiently fast for eigenvalues greater than $k^2$.
Then, $\dot\Gamma_k$ is given by
\be
\dot\Gamma_k=
\frac{1}{2}\Tr\left(\frac{\dot\Delta_k^{(2)}}{\Delta_k^{(2)}}\right)
+\frac{1}{2}\Tr\left(\frac{\dot\Delta_k^{(1)}}{\Delta_k^{(1)}}\right)
+\frac{1}{2}\Tr\left(\frac{\dot\Delta_k^{(0)}}{\Delta_k^{(0)}}\right)
-\Tr\left(\frac{\dot\Delta_{gh,k}^{(1)}}{\Delta_{gh,k}^{(1)}}\right)
-\Tr\left(\frac{\dot\Delta_{gh,k}^{(0)}}{\Delta_{gh,k}^{(0)}}\right)\ .
\label{gammadotexp}
\ee
Note that in the scalar term $\Delta_k^{(0)}$ is a two-by-two matrix,
and the fraction has to be understood as the product of $\dot\Delta_k^{(0)}$
with the inverse of $\Delta_k^{(0)}$. The functional
trace thus involves also a trace over the two-by-two matrix.
Note that any overall prefactor of $\Delta_k$ cancels between
numerator and denominator.

The most convenient choice for the function $R_k$
is the so-called optimized cutoff \cite{optimized}
$R_k(-\bnabla^2)=(k^2+\bnabla^2)\theta(k^2+\bnabla^2)$,
which allows to evaluate the $Q$-integrals in closed form.
In this case the numerator is
$\dot\Delta_k=\dot R_k(-\bnabla^2)=2k^2\theta(k^2+\bnabla^2)$.
Due to the presence of the Heaviside function in the numerator,
in the denominator we can write $\dot P_k(-\bnabla^2)=2k^2$.
The integrations over the eigenvalues of $-\bnabla^2$
that are implicit in the functional traces are
therefore cut off at $k^2$.
The technique that is used to evaluate the functional traces is
explained for example in Appendix A of \cite{Codello:2008vh}.
For the spin-two contribution it gives
\bea
\frac{1}{2}\Tr\left(\frac{\dot\Delta_k^{(2)}}{\Delta_k^{(2)}}\right)
&=&
\frac{1}{2}\frac{1}{(4\pi)^{d/2}}\Bigg[
W(-\bnabla^2,0)\left(Q_{d/2}b_0(\Delta^{(2)})
+Q_{d/2-1}b_2(\Delta^{(2)})
+Q_{d/2-2}b_4(\Delta^{(2)})\right)
\nonumber\\
&&
\qquad\qquad
+W'(-\bnabla^2,0)\bR\left(Q_{d/2}b_0(\Delta^{(2)})
+Q_{d/2-1}b_2(\Delta^{(2)})\right)
\nonumber\\
&&
\qquad\qquad
+\frac{1}{2}W''(-\bnabla^2,0)\bR^2\left(Q_{d/2}b_0(\Delta^{(2)})\right)
+\ldots\Bigg],
\label{euterpe}
\eea
where
$W(-\bnabla^2,\bR)=\frac{\dot\Delta_k^{(2)}}{\Delta_k^{(2)}}$
and primes denote derivatives with respect to $\bR$.
The coefficients $Q_n$ and the heat kernel coefficients
$b_n$ are listed in Appendix A.
Similar formulas hold for the spin one and spin zero
sectors and for the ghosts.
With these data one can write the expansion of (\ref{gammadotexp})
in powers of $\bar R$ and comparing with (\ref{gammadotabc}) one can
read off the coefficients $A_1$, $B_1$ and $C_1$.

\section{Results}
In the following we shall discuss mainly the case $d=4$,
but we will point out some results that hold in any dimension.

We begin with the coefficient $A_1$, that is the simplest of the three.
Normally the vacuum energy, which diverges like $k^d$,
is simply proportional to the number of degrees of freedom.
As pointed out in \cite{falls1}, for pure gravity this is $d(d-3)/2$.
The general result is actually more complicated, but it
reduces to the expected value if one assumes either $\tilde\Lambda=0$,
or $\omega=1/2$ and $b\to\pm\infty$.
We will discuss the meaning of this second choice later.
In both cases one has
\be
A_1=\frac{16\pi(d-3)}{(4\pi)^{d/2}\Gamma(d/2)} ,
\ee
independently of the choice of gauge and parametrization.
In particular, note that in three dimensions $A_1=0$,
reflecting the absence of propagating gravitons,
and in four dimensions $A_1=1/\pi$.
We will not discuss the coefficient $A_1$ anymore.

For $\tilde\Lambda=0$, also $B_1$ and $C_1$ simplify considerably.
Unless otherwise stated, we will therefore consider
only this case in what follows\footnote{
While this may seem a strong restriction, we shall see that there
are choices of gauge and parametrization where the results are
automatically $\tilde\Lambda$-independent.}.
The coefficients $B_1$ and $C_1$ then depend on the four parameters
$m$, $\omega$, $a$ and $b$.
One can get some understanding of the behavior of these functions,
by fixing either the parametrization or the gauge,
and studying the dependence on the remaining two parameters.
In Figs.~(\ref{fig.linpar}) and (\ref{fig.exppar})
we fix the linear parametrization $\omega=0$, $m=0$
or the exponential parametrization $\omega=1/2$, $m=0$
and plot $B_1$ and $C_1$ as functions of the gauge parameters
$a$ and $b$.
In Figs.~(\ref{fig.fdd}) and (\ref{fig.phys})
we fix the Feynman-de Donder gauge $a=1$, $b=1$
or the physical gauge $a=0$, $b\to\pm\infty$
and plot $B_1$ and $C_1$ as functions of $m$ and $\omega$.
Other choices such as the Landau-de Donder gauge
$a=0$, $b=1$ or the ``tracefree'' conditions
$b=0$ yield similar pictures.
We now discuss some remarkable special cases.

\subsection{Fixing $\omega=0$}

In the standard linear parametrization ($m=\omega=0$) in $d=4$ we have
\bea
B_1&=&\frac{a  \left(-6 b ^2+36 b -62\right)-3 \left(7 b ^2-50 b +79\right)}{8 \pi  (b -3)^2}\ ,
\label{B1lin}
\\
C_1&=&\frac{1}{17280\pi^2(b-3)^4}
\Big[135a^2\left(3b^4-36b^3+162b^2-324b+259\right)
\nonumber\\
&&
\qquad\qquad\qquad\qquad
-180a\left(3b^4-36b^3+176b^2-360b+297\right)
\nonumber\\
&&
\qquad\qquad\qquad\qquad
+4\left(431b^4-3822b^3+14904b^2-26298b+17901\right)\Big]\ .
\label{C1lin}
\eea
We show in Appendix B that the formula for $C_1$ 
is in agreement with an old calculation
of Kallosh et al. \cite{kallosh}.
We also compare our results for $\omega=0$ and $\omega=1$,
but general $m$, with the calculations in reference \cite{Kalmykov:1995fd}.
To the extent that the calculations overlap,
they are again seen to agree.

\begin{figure}
\begin{center}
\resizebox{1.0\columnwidth}{!}
{\includegraphics*{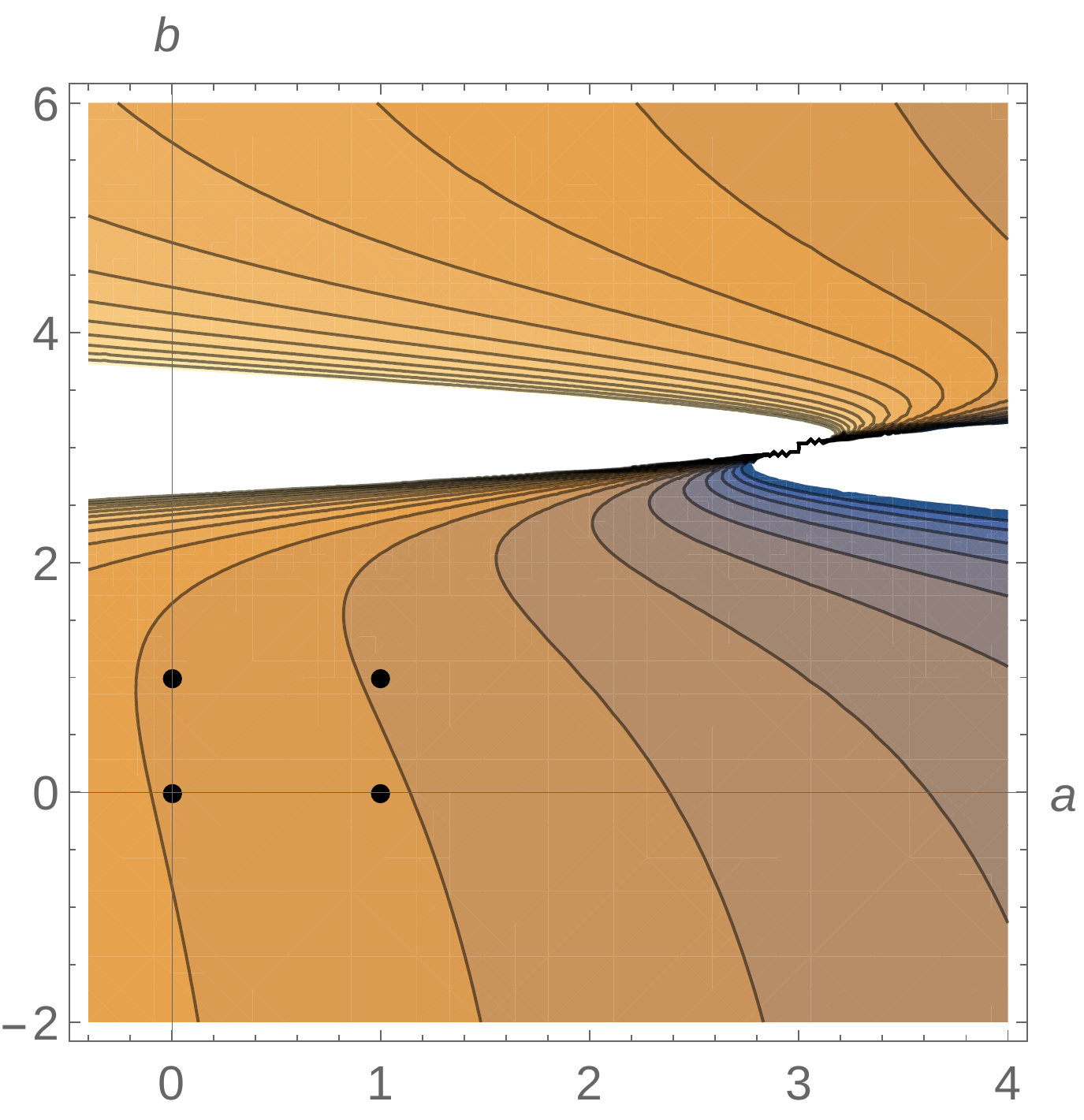}\qquad
\includegraphics*{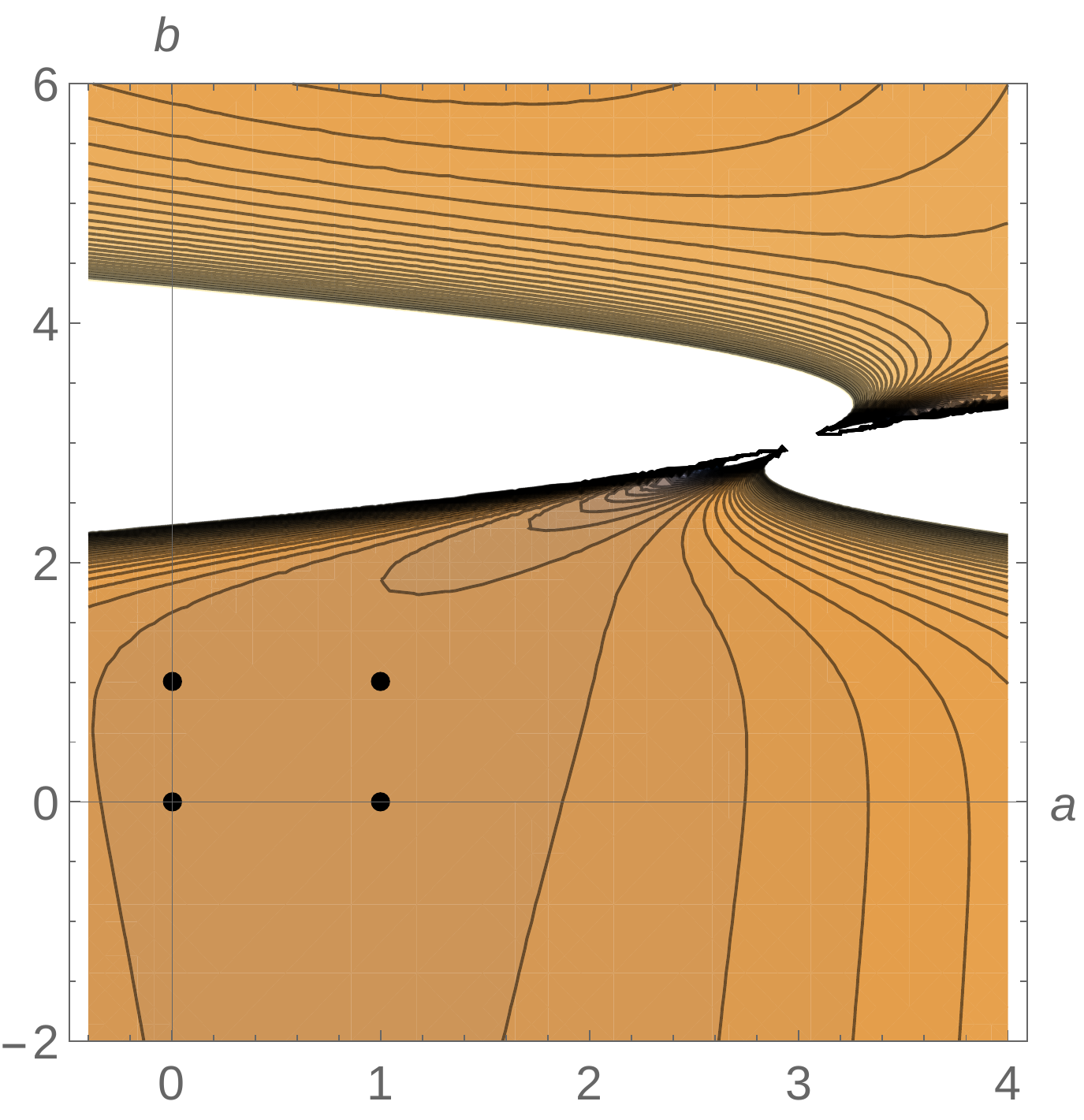}}
\\
\resizebox{0.8\columnwidth}{!}
{\includegraphics{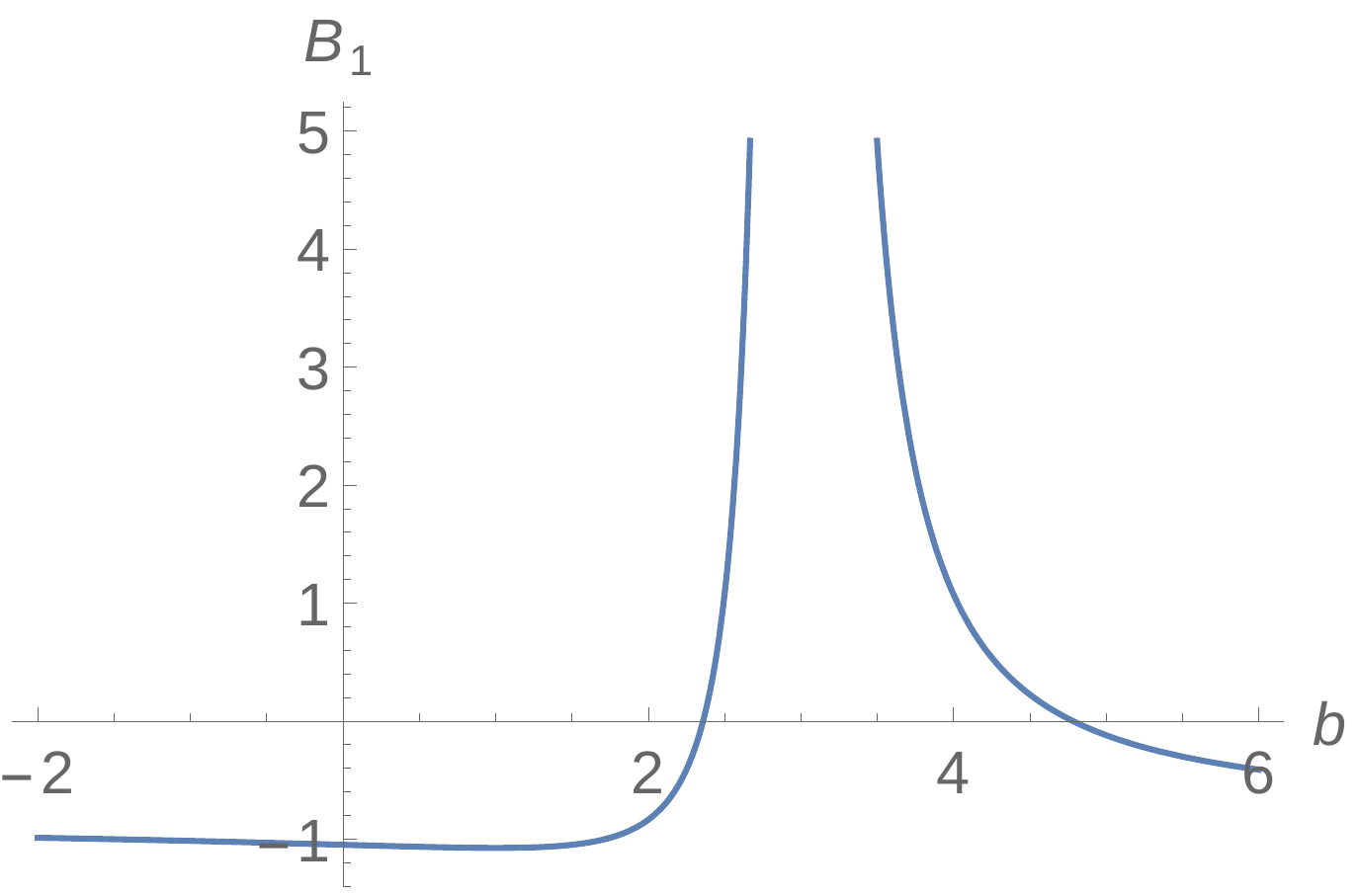}\hspace{30mm}
\includegraphics{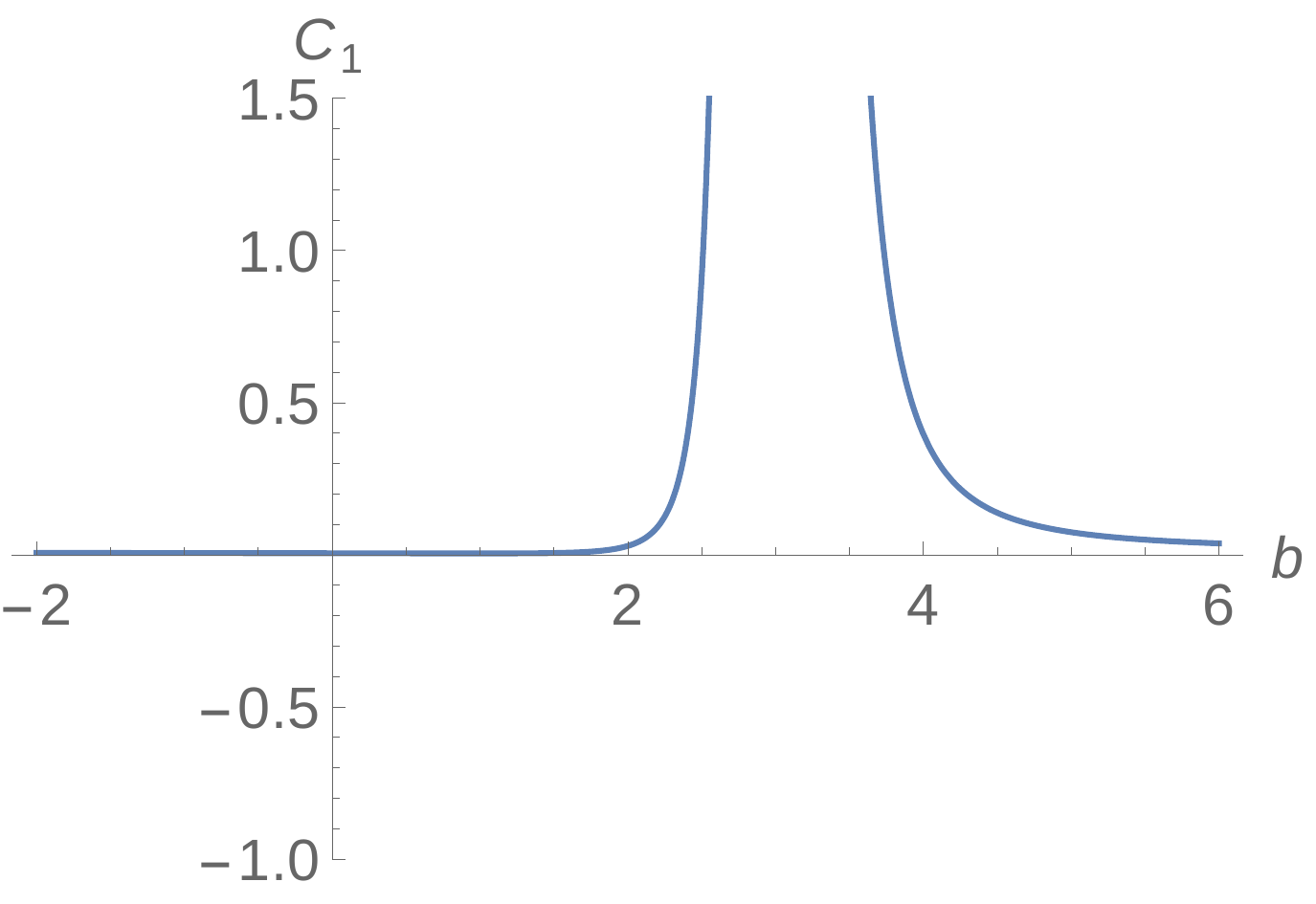}}
\caption{The coefficients $B_1$ (left) and $C_1$ (right)
in $d=4$ for $\tilde\Lambda=0$,
in the standard linear parametrization $\omega=0$, $m=0$.
The functions decrease going from lighter to darker tones.
The zero-level lines of $B_1$ are the ones ending on the left
near $b=5$ and $b=2.3$. The interval between level lines is $0.34$.
The function $B_1$ goes to $+\infty$ on the line $b=3$,
left of the point $(3,3)$ and to $-\infty$ right of that point.
The zero-level line of $C_1$ is the biggest of the loops that
are seen emanating leftwards from the singular point in $(3,3)$,
and a similar loop on the opposite side.
$C_1$ is positive outside the loop and goes to $+\infty$
on the line $b=3$.
The interval between level lines is 0.068.
The plots in the lower row are cuts through the line $a=0$.
The four dots mark the familiar gauges $a=0,1$, $b=0,1$.}
\label{fig.linpar}
\end{center}
\end{figure}

The functions (\ref{B1lin}) and (\ref{C1lin}) are plotted
in Fig.~(\ref{fig.linpar}).
One sees that there is a divergence on the line $b=3$.
This can be attributed to the failure of the gauge condition at $b=3$.
Elsewhere, the gauge dependence is relatively weak.
As mentioned in the Introduction, the most reliable results
are obtained for $a\to 0$.
One normally considers the gauges where $b=1$ or $b=0$,
which are indicated by black dots, but there is no reason to discard
large values of $b$, in particular for $b\to\pm\infty$ we have
\bea
B_1&=&-\frac{3(7+2a)}{8\pi}\ ,
\\
C_1&=&-\frac{1724-540a+405a^2}{17280\pi^2}\ .
\nonumber
\eea
Taking the limit $b\to\pm\infty$ corresponds to imposing the
condition $h=0$ strongly.
In the limit $a\to 0$ one also imposes $\hat\xi_\mu=0$.
On the other hand for $a\to 0$ and $b\to 0$
one imposes $\hat\xi_\mu=0$ and $\hat\sigma=0$ strongly.
These were called ``physical gauges'' in \cite{pv1}.

In order to get a feeling of the numerical variability,
the following table gives the values of the coefficients for some
selected gauges
\begin{center}
\begin{tabular}{|c||c|c|}
\hline
\rule[-2mm]{0mm}{6mm}
Gauge &  $B_1$ & $C_1$\\
\hline
\hline
\rule[-1mm]{0mm}{5mm}
$a=0$, $b=0$ &  $-1.05$   & 0.0052 \\
\hline
\rule[-1mm]{0mm}{5mm}
$a=0$, $b=1$ &  $-1.07$  & 0.0046 \\
\hline
\rule[-1mm]{0mm}{5mm}
$a=0$, $b=\pm\infty$ &  $-0.83$  & 0.010 \\
\hline
\rule[-1mm]{0mm}{5mm}
$a=1$, $b=1$ &  $-1.39$  &  0.0025 \\
\hline
\end{tabular}
\end{center}

\subsection{Fixing $\omega=1/2$}

In the exponential parametrization a simplification occurs:
$B_1$ and $C_1$ become independent of $m$.
In four dimensions one has
\bea
B_1&=&-\frac{159-8a-90b+15b^2}{8\pi(b-3)^2} \ ,
\label{enea}
\\
C_1&=&-\frac{55971-2160a^2-68148b
+29754b^2
-6852b^3+571b^4
-360a(9-18b+b^2)}{17280(b-3)^4\pi^2}\ .
\nonumber
\eea
The independence on $m$ can be understood by considering the
Hessian (\ref{ehgf1}).
For $\omega=1/2$ the spin-one and spin-two operators
are independent of $m$, while in the determinant of the
scalar sector $m$ only appears in an overall prefactor.

\begin{figure}
\begin{center}
\resizebox{1.0\columnwidth}{!}
{\includegraphics{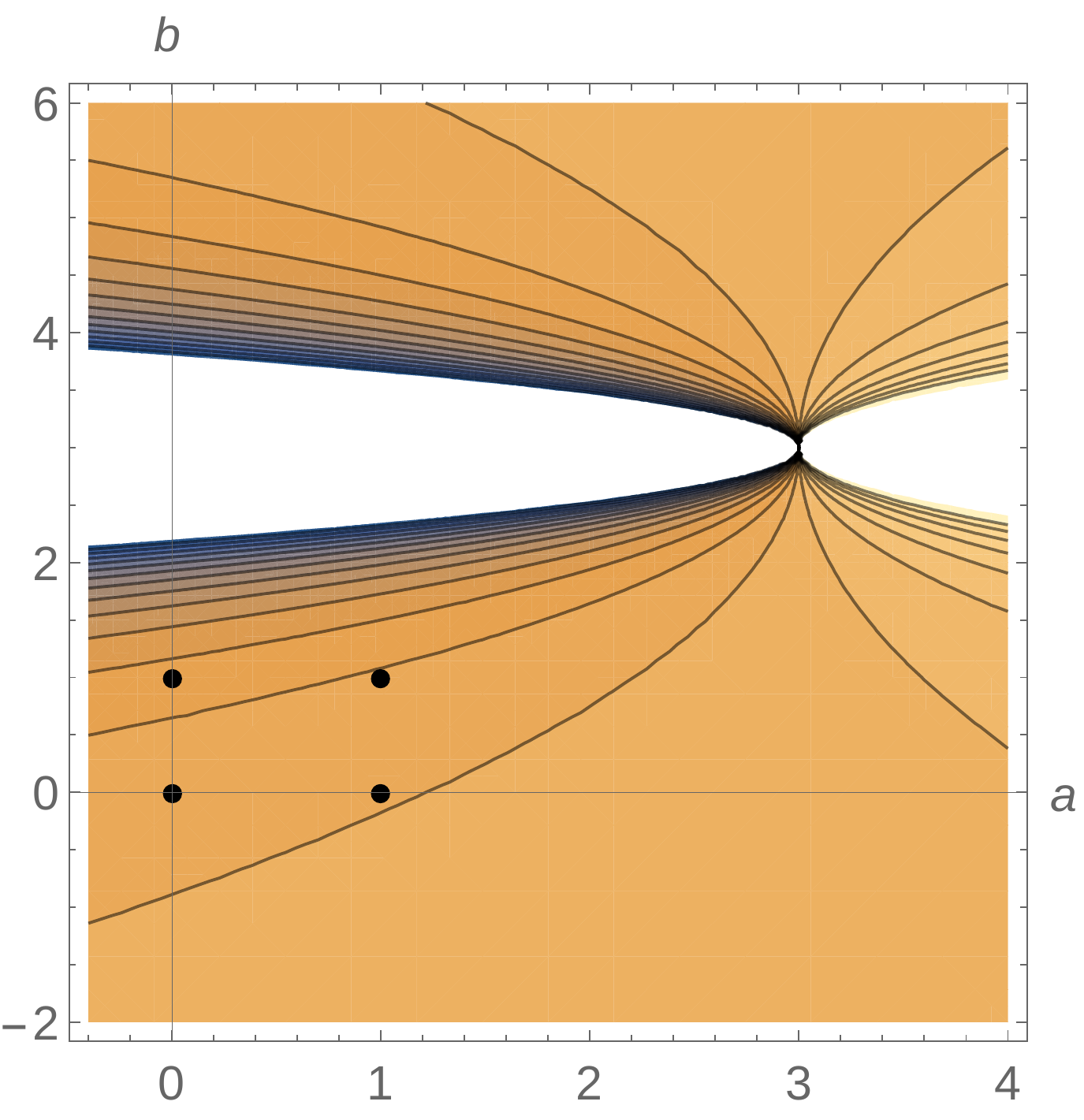}\qquad
\includegraphics{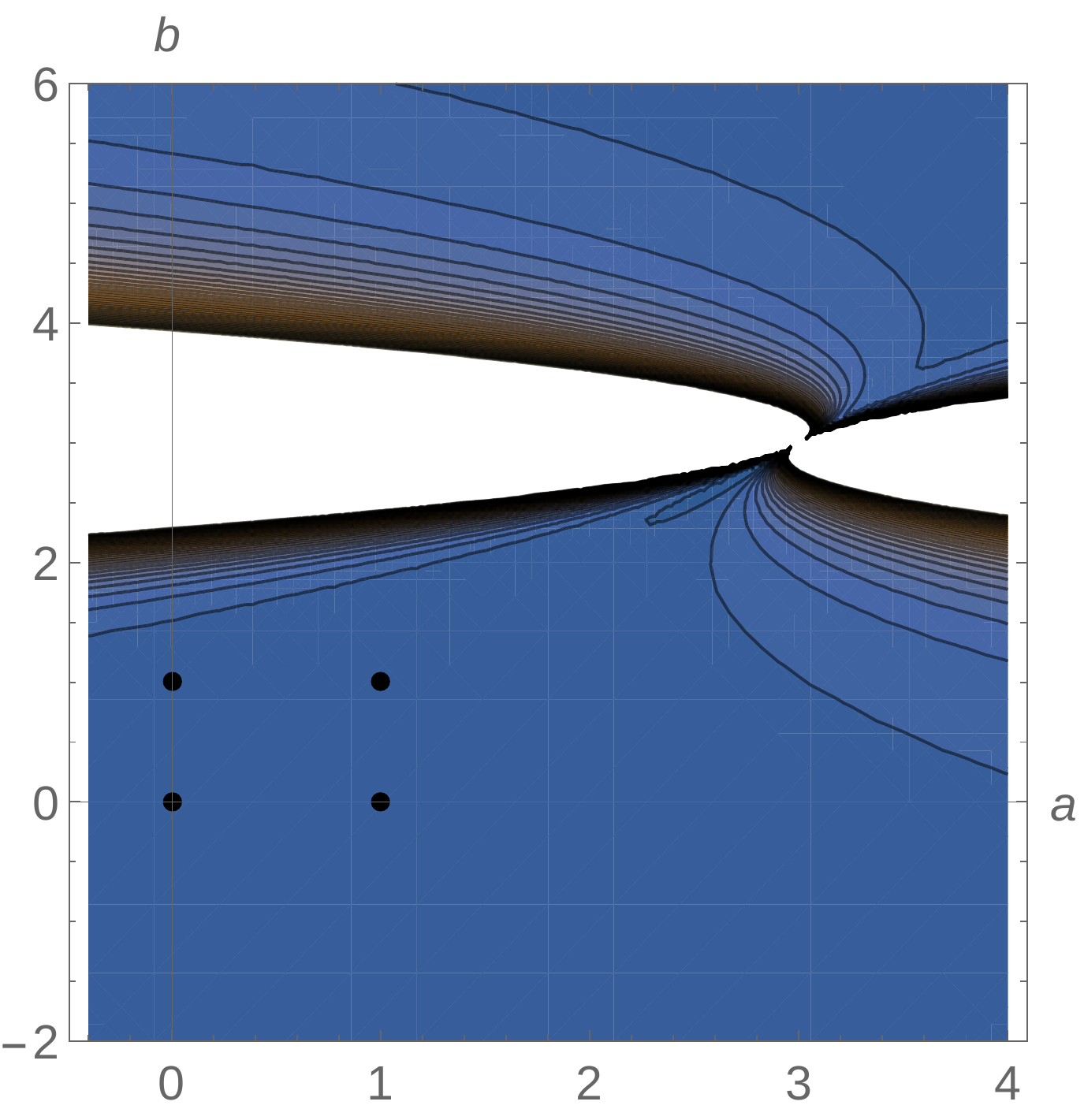}}
\\
\resizebox{0.80\columnwidth}{!}
{\includegraphics{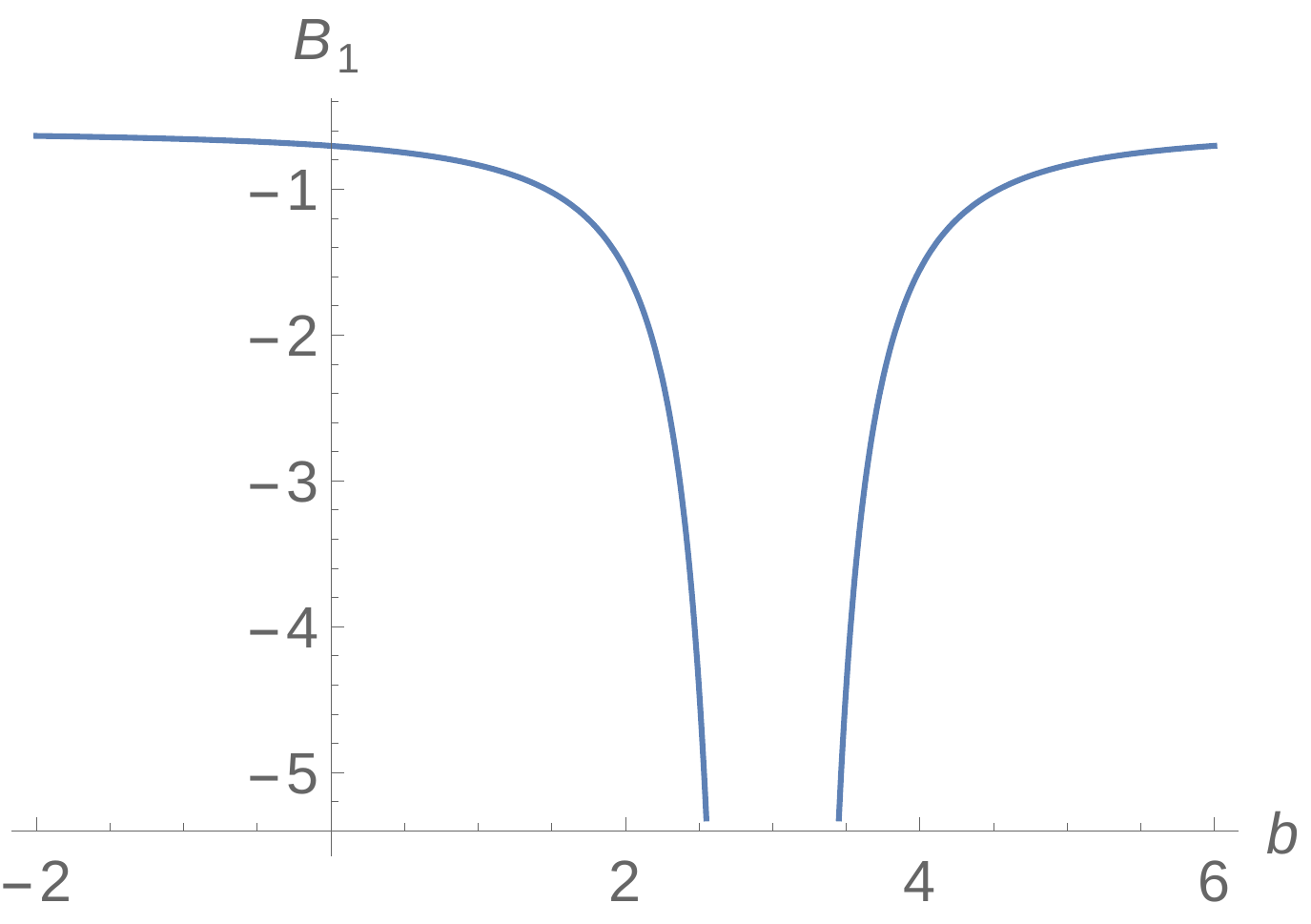}\hspace{30mm}
\includegraphics{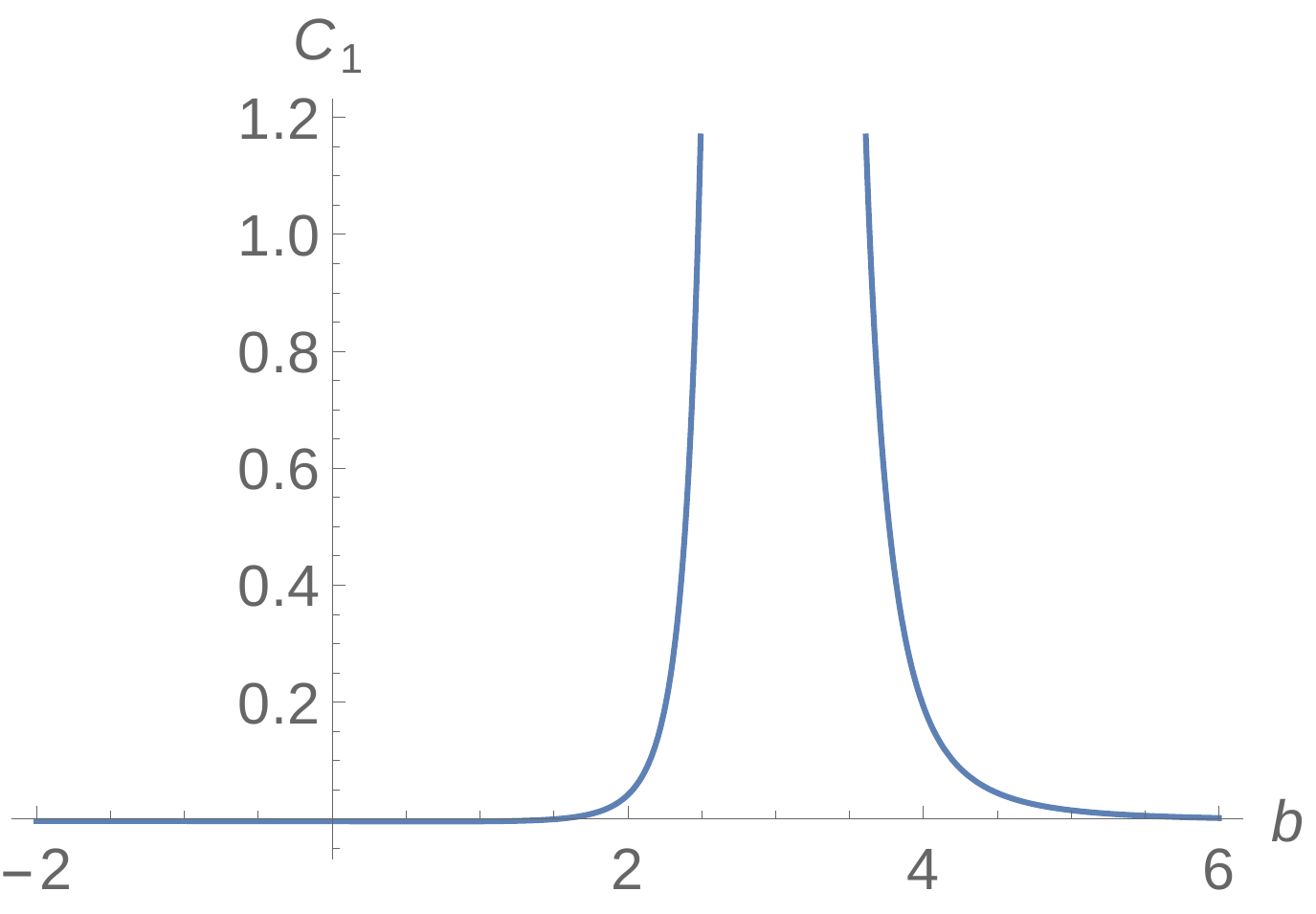}}
\caption{The coefficients $B_1$ (left) and $C_1$ (right)
in $d=4$ for $\tilde\Lambda=0$,
in the exponential parametrization $\omega=1/2$.
The function $B_1$ is negative throughout most of the plot,
with the zero-level line being the the second innermost parabola
ending on the right a little below $b=4$ and above $b=2$.
The interval between level lines is $0.11$.
In contrast to the linear parametrization,
the function $B_1$ goes to $-\infty$ on the line $b=3$,
left of the point $(3,3)$ and to $+\infty$ right of that point.
The function $C_1$ is slightly negative in the areas in the top right
and bottom of the figure with the zero-level lines being the outermost
lines both in the upper and lower regions,
and goes to $+\infty$ on the line $b=3$.
The plots in the lower row are cuts through the line $a=0$.
The four dots mark the familiar gauges $a=0,1$, $b=0,1$.
Both coefficients diverge at $b=3$
and become independent of $a$ for $b\rightarrow\pm\infty$.}
\label{fig.exppar}
\end{center}
\end{figure}

The functions (\ref{enea}) are plotted in Fig.~(\ref{fig.exppar}).
The divergence for $b=3$ is still present
but elsewhere the gauge-dependence is again weak.
Numerical values are given in the following table:
\begin{center}
\begin{tabular}{|c||c|c|}
\hline
\rule[-2mm]{0mm}{6mm}
Gauge &  $B_1$ & $C_1$\\
\hline
\hline
\rule[-1mm]{0mm}{5mm}
$a=0$, $b=0$ &  $-0.70$   & $-0.0041$ \\
\hline
\rule[-1mm]{0mm}{5mm}
$a=0$, $b=1$ &  $-0.84$  &  $-0.0041$ \\
\hline
\rule[-1mm]{0mm}{5mm}
$a=0$, $b=\pm\infty$ &  $-0.60$  & $-0.0033$ \\
\hline
\rule[-1mm]{0mm}{5mm}
$a=1$, $b=1$ & $-0.76$  & $-0.0044$  \\
\hline
\end{tabular}
\end{center}
In the limit $b\to\pm\infty$ a further simplification occurs:
the dependence on $a$ automatically disappears:
\be
A_1=\frac{1}{\pi}\ \ ,\qquad
B_1=-\frac{15}{8\pi}\ \ ,\qquad
C_1=-\frac{571}{17280 \pi ^2}\ .
\label{bcphys}
\ee
In fact, a stronger statement can be made:
if one chooses the exponential parametrization
and the partial gauge condition $b\to\pm\infty$
the coefficients $B_1$ and $C_1$ become automatically independent
of $\tilde\Lambda$, $m$ and $a$, in any dimension:
\bea
B_1&=&\frac{d^5-4 d^4-9 d^3-48 d^2+60 d+24}
{(4\pi)^{d/2-1}3(d-1)d^2\Gamma\left(\frac{d}{2}\right)}\ ,
\\
C_1&=&\frac{5 d^8-37 d^7-17 d^6-743 d^5+1668 d^4+684 d^3+16440 d^2-13680 d-8640}
{(4\pi)^{d/2}1440(d-1)^2 d^3 \Gamma \left(\frac{d}{2}\right)}\ .
	\nonumber
\eea
The reason for the independence on $\tilde\Lambda$ can be understood
as follows \cite{pv1}.
In exponential parametrization the expansion of the cosmological
term contains only terms proportional to $h$,
the trace of the fluctuation field.
The gauge $b\to\pm\infty$ amounts to imposing $h=0$,
so the cosmological term disappears from the effective action.
In contrast, in the linear parametrization, the second order expansion
of the cosmological term contains a term proportional to
$h_{\mu\nu}h^{\mu\nu}$.
This contributes a ``mass'' term proportional to $\Lambda$
to the graviton propagator,
which then gives rise to denominators of the form $1-2\tilde\Lambda$
that appear everywhere in $\dot\Gamma_k$.

\subsection{Fixing the gauge}

\begin{figure}
\begin{center}
\resizebox{1.0\columnwidth}{!}
{\includegraphics{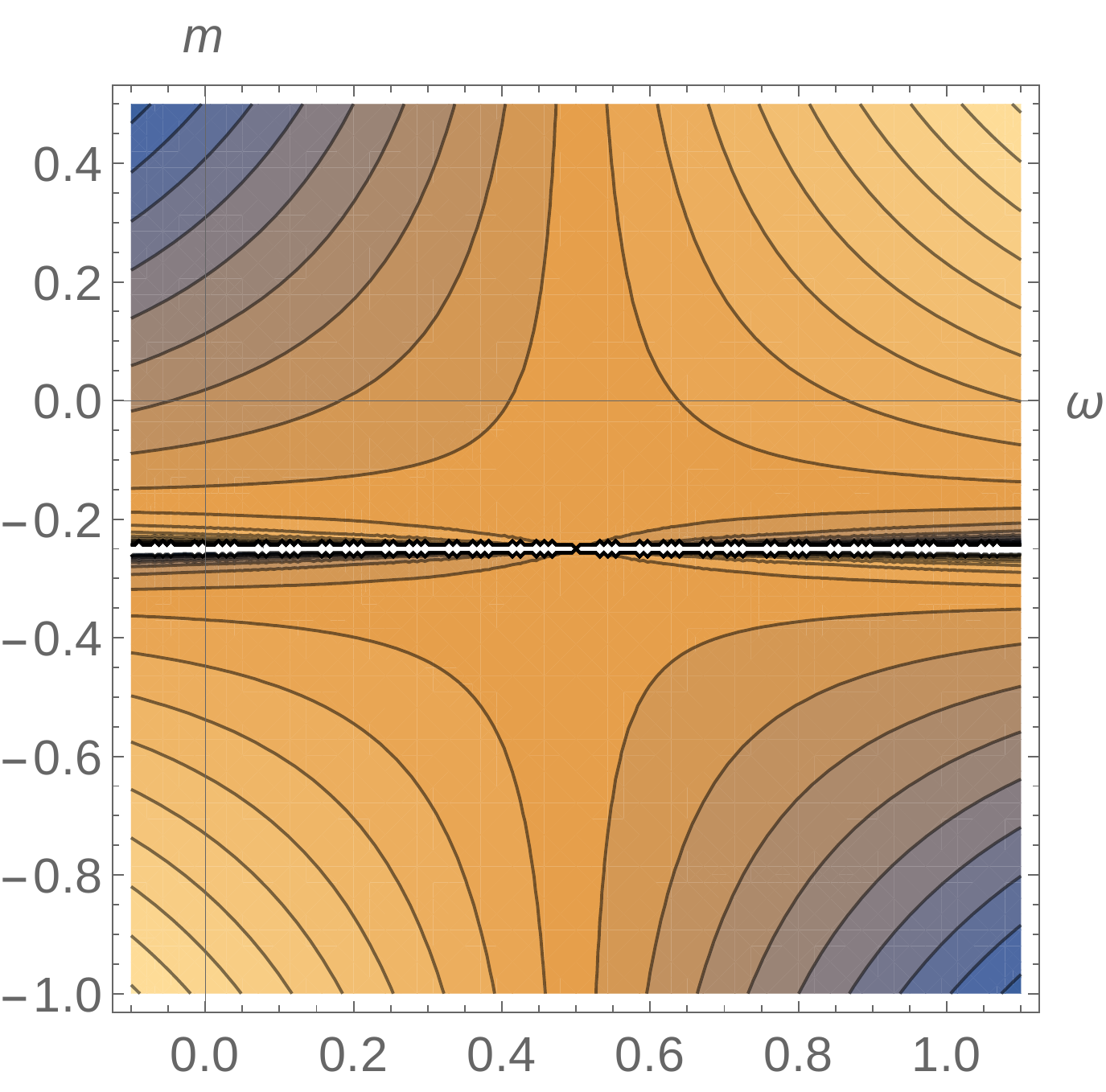}\qquad\qquad
\includegraphics{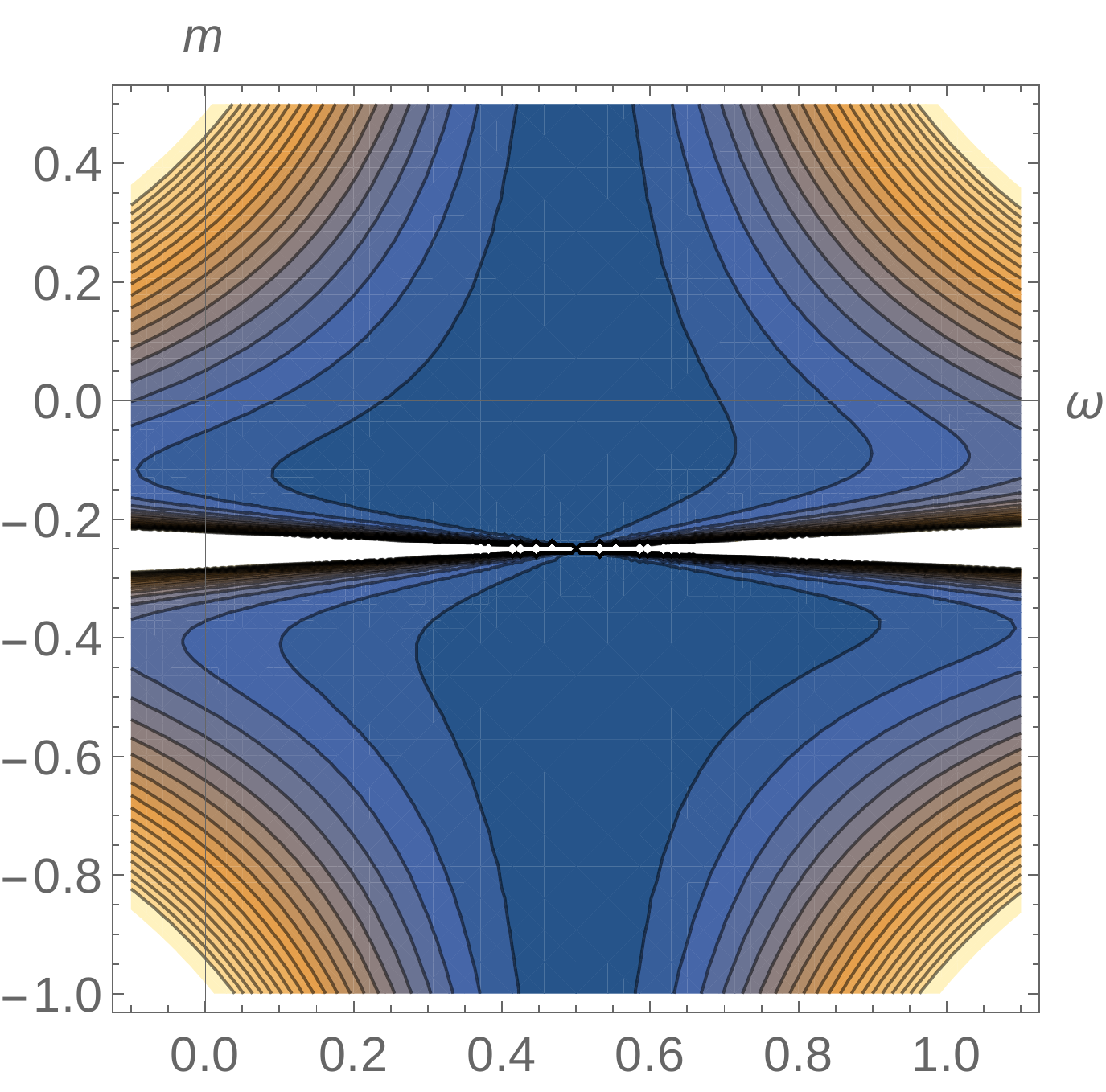}}
\\
\resizebox{0.80\columnwidth}{!}
{\includegraphics{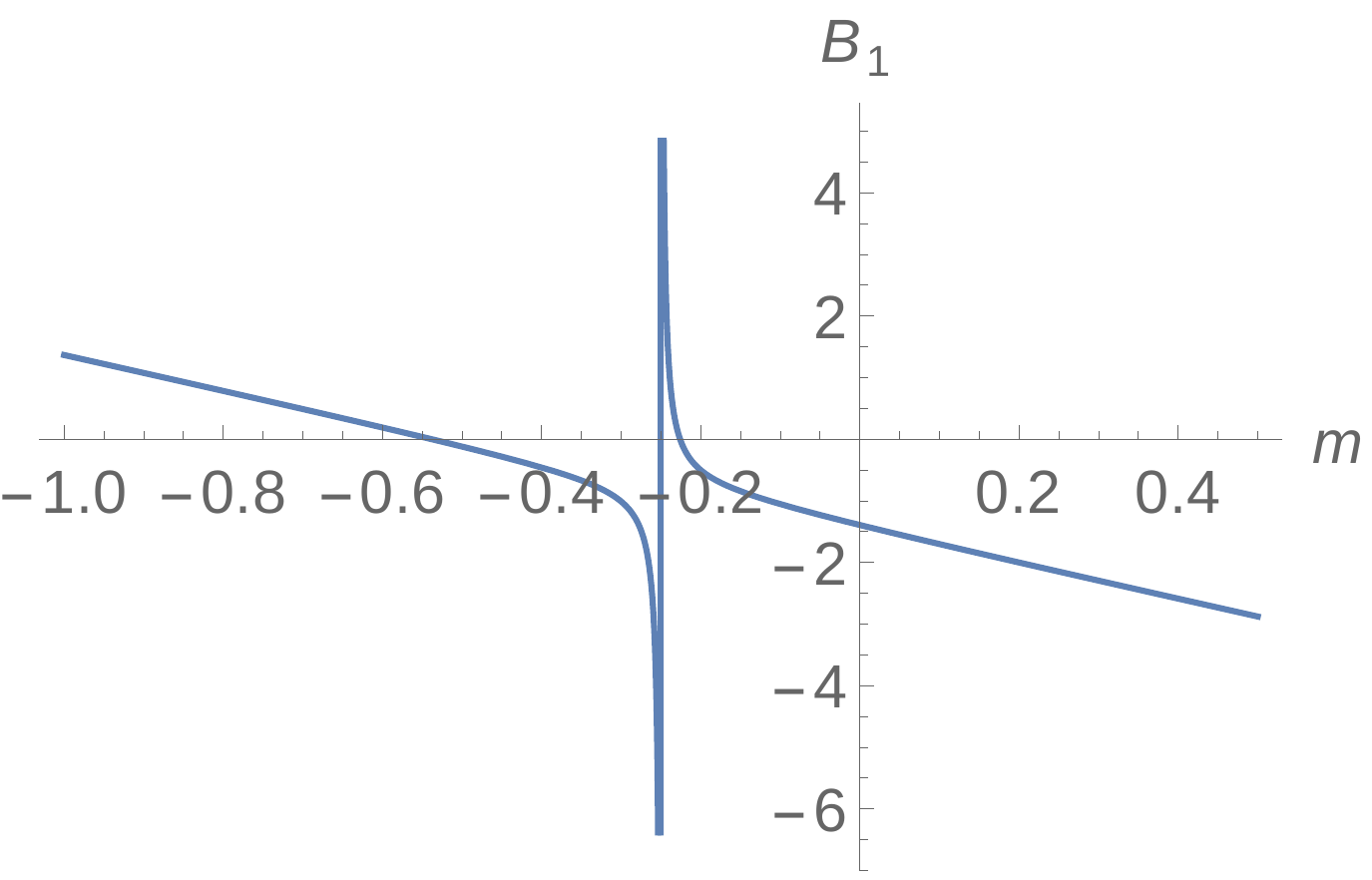}\hspace{30mm}
\includegraphics{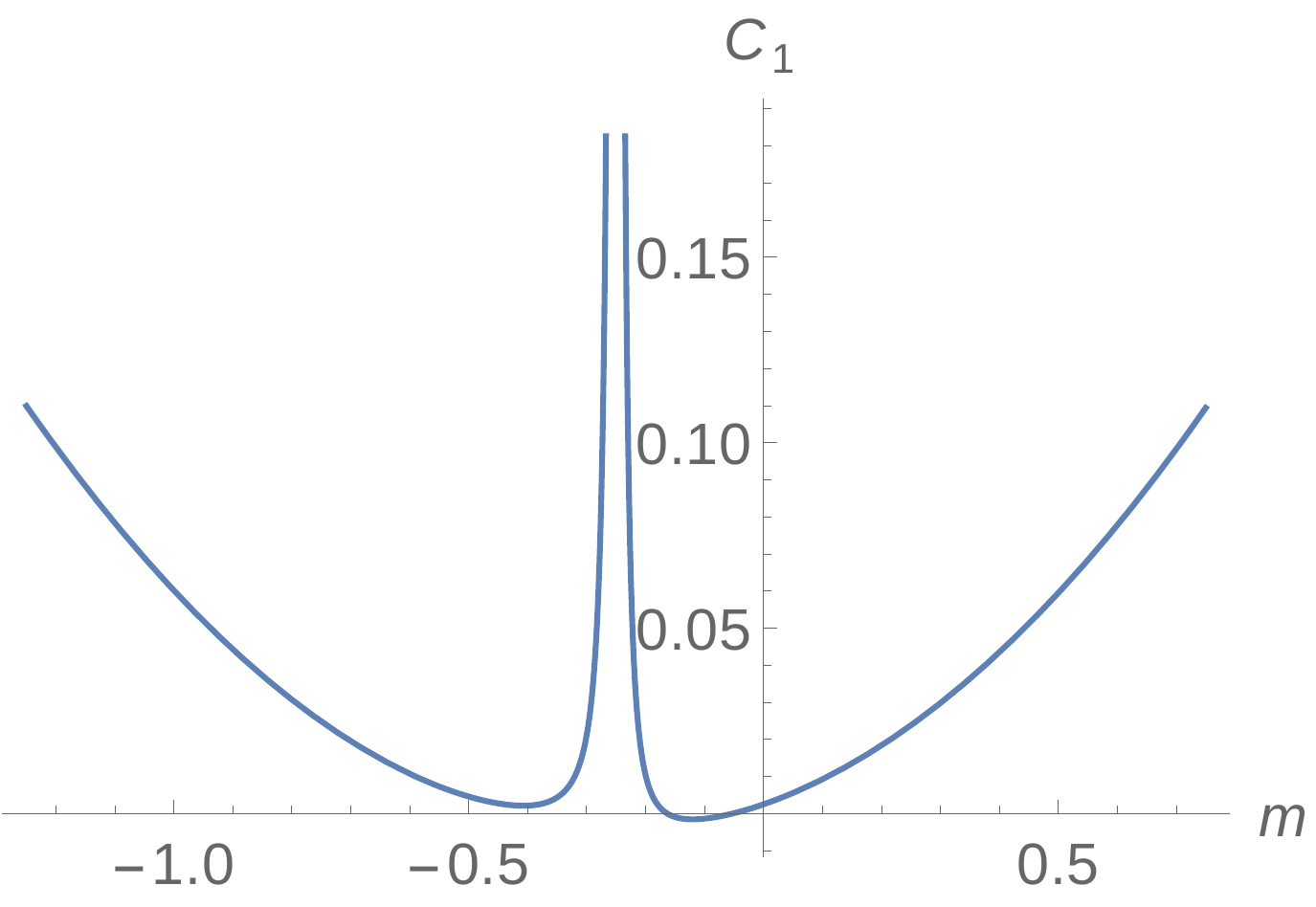}}
\caption{The coefficients $B_1$ (left) and $C_1$ (right)
in $d=4$ for $\tilde\Lambda=0$,
in the Feynman-de Donder gauge $a=1$, $b=1$,
as functions of $\omega$ and $m$.
$B_1$ is positive in the lower left and upper right corners.
The zero-level lines are the ones ending near $m=-0.5$
on the left and $m=0$ on the right.
$C_1$ is negative in the two darkest regions in the center
of the plot.
The structure of the singularity at $m=-1/4$ can be
understood from the plots in the lower row,
which are cuts through the line $\omega=0$.
Both coefficients are constant on the line $\omega=1/2$.
}
\label{fig.fdd}
\end{center}
\end{figure}

An alternative way to cut up the four-parameter space is to
fix the gauge.
One can then study the dependence of the off-shell effective action
on the choice of the measure.
Figure (\ref{fig.fdd}) shows the coefficients $B_1$ and $C_1$
as functions of $\omega$ and $m$ in the familiar Feynman-de Donder gauge.
In interpreting these figures one has to recall that
the vertical lines at $\omega=0$ and $\omega=1$
correspond to treating the (densitized) or inverse metric
as fundamental variables, while the line $\omega=1/2$
corresponds to the exponential parametrization.
Intermediate values do not have direct physical interpretation.
The vertical axis measures the weight of the quantum field.
Near $\omega=0$ and $\omega=1$ the functions $B_1$ and $C_1$
are approximately linear and quadratic, respectively,
apart from a singularity at $m=-1/4$.
The singularity is located precisely where the relation between
$g_{\mu\nu}$ and $\gamma_{\mu\nu}$ is not invertible.
Exactly at $\omega=1/2$, $B_1$ and $C_1$ are both constant:
\be
B_1=-\frac{19}{8\pi}\ \ ,\qquad
C_1=-\frac{751}{17280 \pi ^2}\ .
\ee
Slightly different gauge choices, for example the Landau-de Donder
gauge $b=1$, $a=0$, or ``traceless'' gauges\footnote{By this we mean that the gauge condition is imposed
on the traceless part of $h_{\mu\nu}$, not that the trace
$h_{\mu\nu}$ is zero. Such gauge choice is obtained in the
opposite limit $b\to\pm\infty$.}
$b=0$, give qualitatively the same results.

\begin{figure}
\begin{center}
\resizebox{1.0\columnwidth}{!}
{\includegraphics{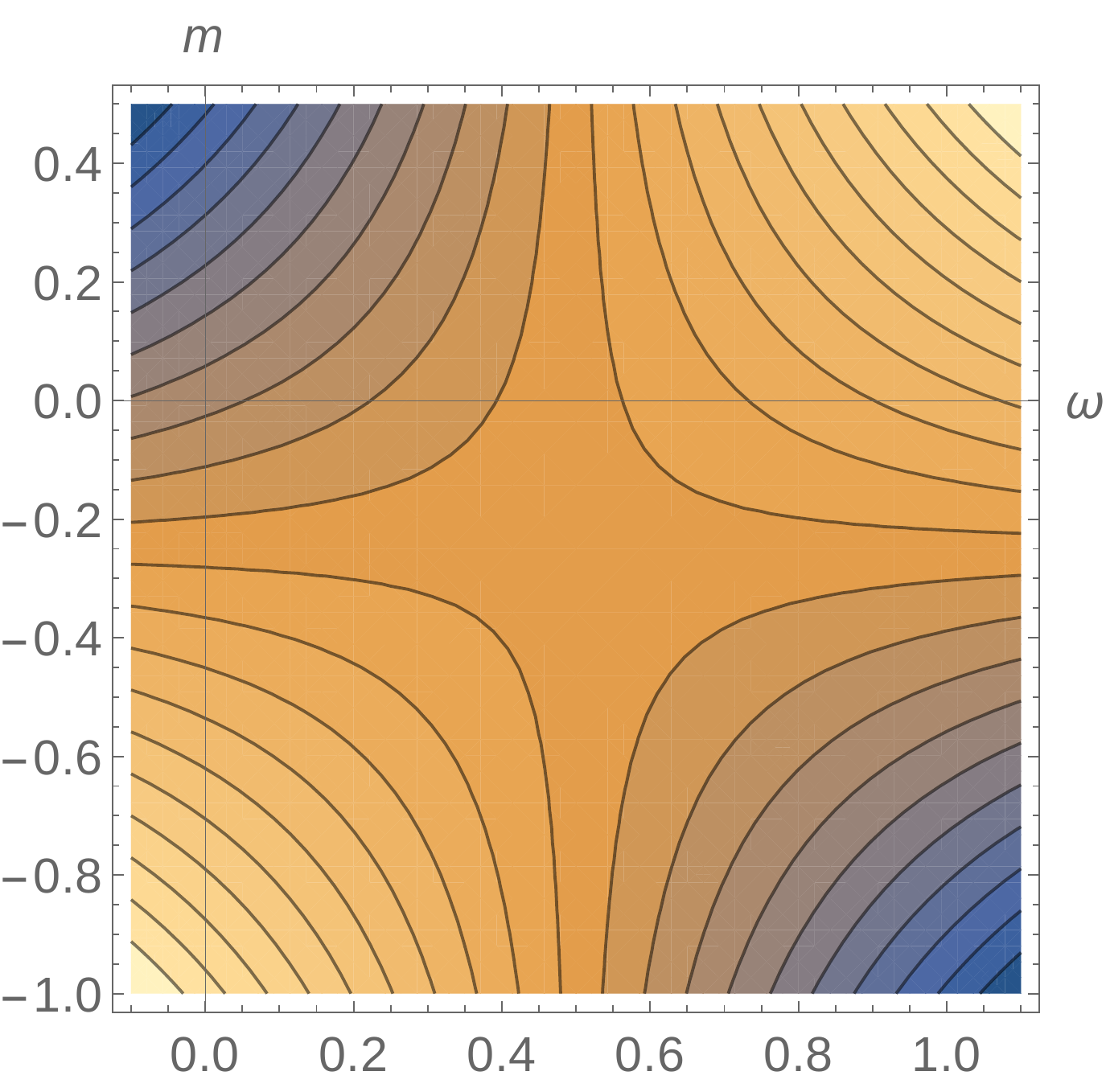}\qquad
\includegraphics{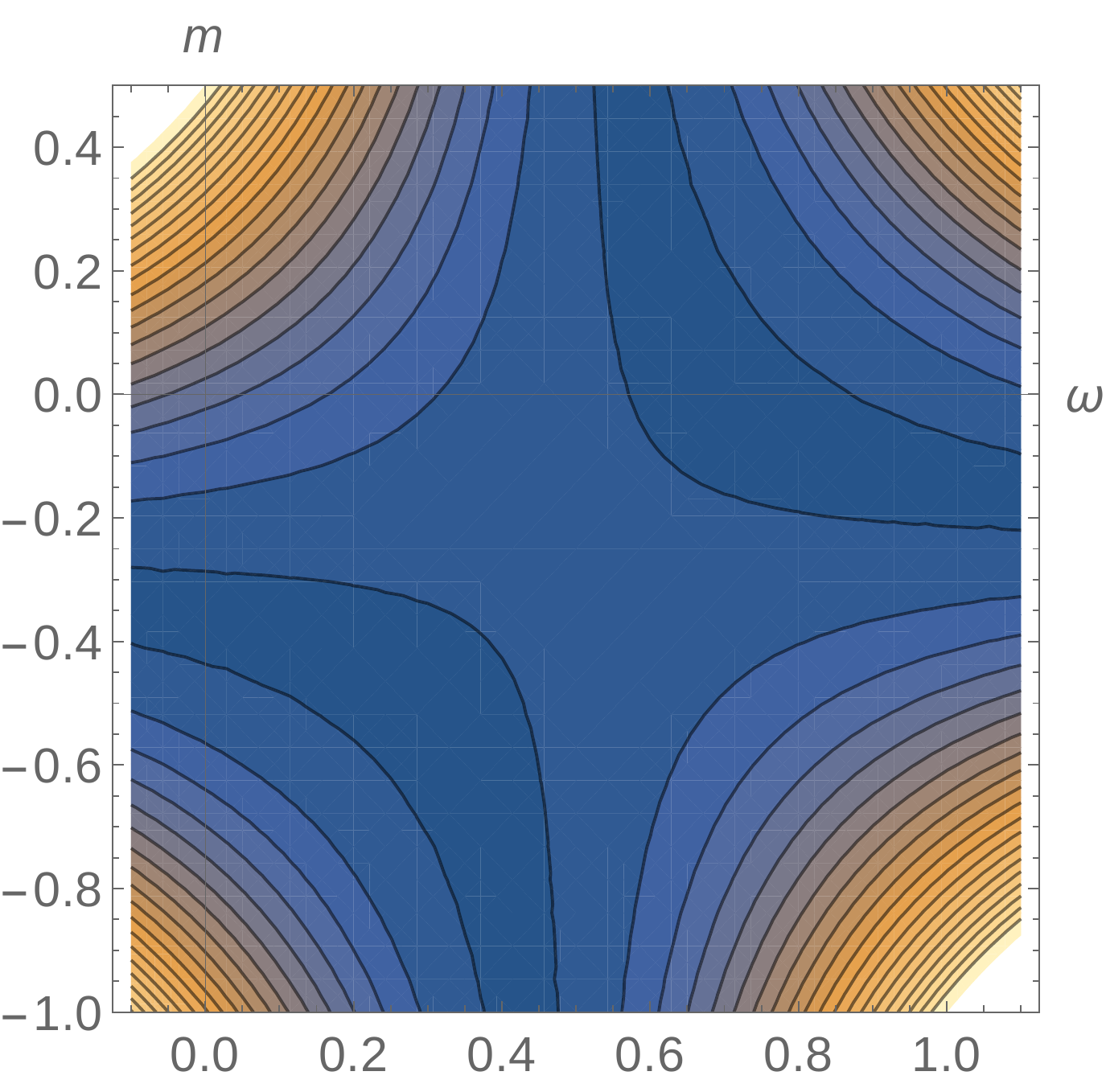}}
\\
\resizebox{0.80\columnwidth}{!}
{\includegraphics{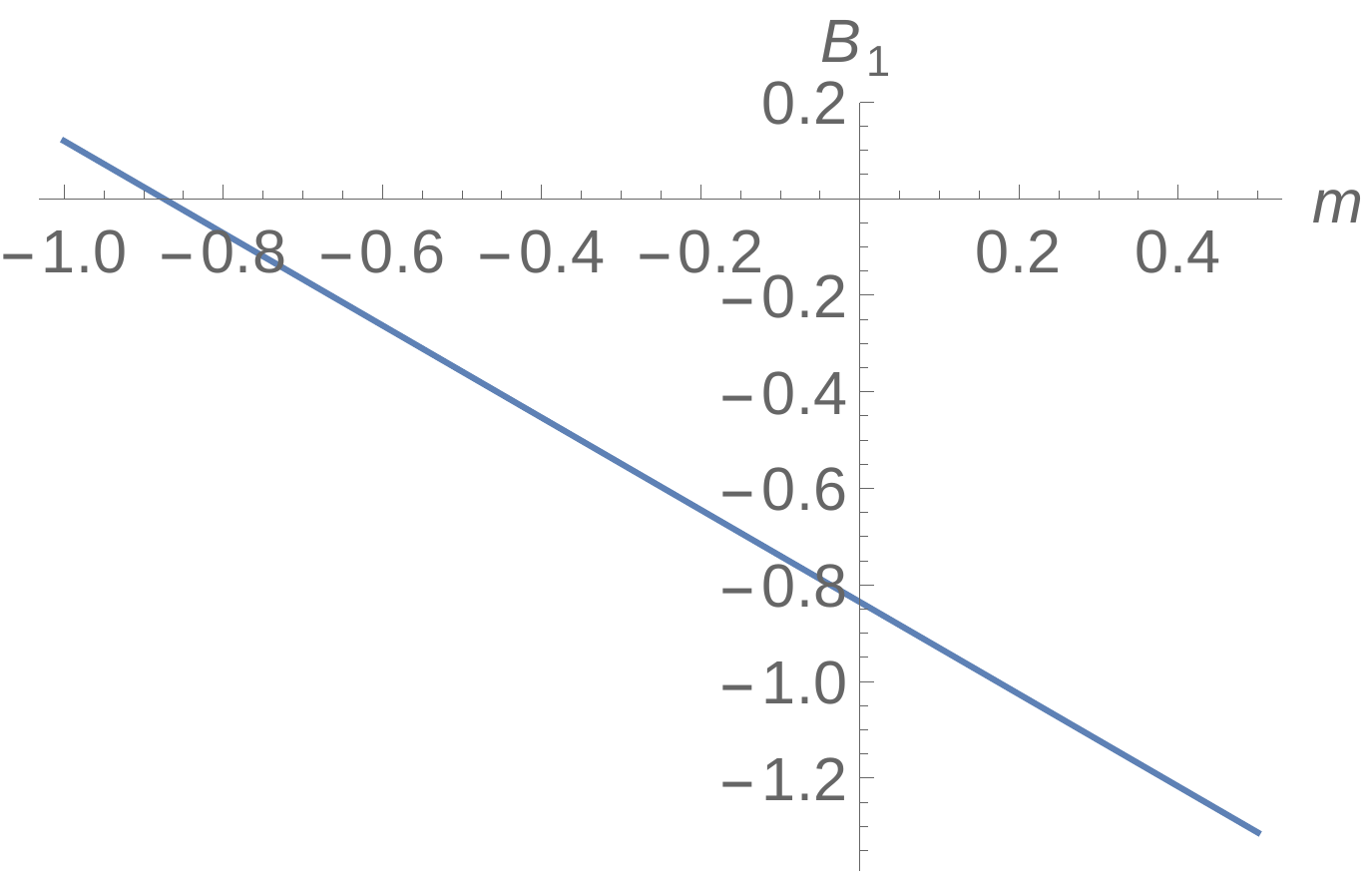}\hspace{30mm}
\includegraphics{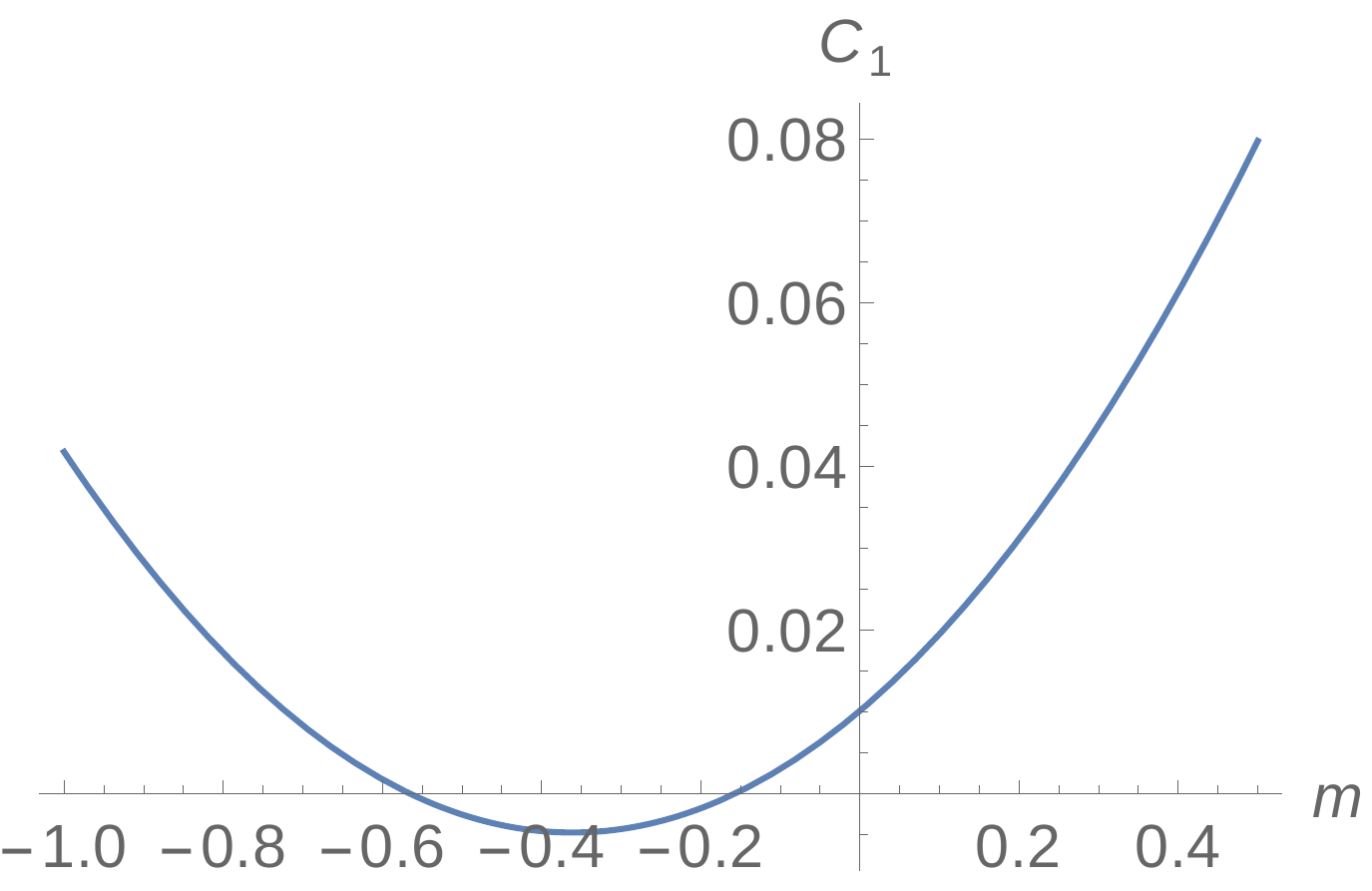}}
\caption{The coefficients $B_1$ (left) and $C_1$ (right)
in $d=4$ for $\tilde\Lambda=0$,
in the ``unimodular physical'' gauge $a=0$, $b\to\pm\infty$,
as functions of $\omega$ and $m$.
The color code is as in the previous figure.
The zero-level lines of $B_1$ now end near $m=-0.7$ on the left and
$0.2$ on the right.
$C_1$ is negative in the three darkest regions (aligned north-east and south-west) in the center.
The simple plots below each contour plot are cuts through the
line $\omega=0$.
In this case there are no divergences at $m=-1/4$:
$B_1$ is simply linear and $C_1$ is simply quadratic.
Both coefficients are constant on the lines $\omega=1/2$
and $m=-1/4$.
}
\label{fig.phys}
\end{center}
\end{figure}

Finally, Fig.~(\ref{fig.phys}) shows the coefficients $B_1$ and $C_1$
as functions of $\omega$ and $m$ in the ``physical'' gauge
$a=0$, $b\to\pm\infty$, which corresponds to imposing
$h=0$, $\xi_\mu=0$.
The general behavior is very similar to that of Fig.~(\ref{fig.fdd})
with the striking difference that the singularity at $m=-1/4$ is absent.
On the line $\omega=1/2$ the coefficients are given by (\ref{bcphys}).
On the line $m=-1/4$ (where the present calculation is not
supposed to be valid) one also obtains the same values.


\goodbreak

\section{Duality}
Figures (\ref{fig.fdd}) and (\ref{fig.phys}) have a reflection symmetry
about the point with coordinates $\omega=1/2$, $m=-1/4$.
This is a special case of a much more general relation:
in any dimension, for any value of $\tilde\Lambda$ and in any gauge,
the functions $B_1$ and $C_1$ have the following property:
\bea
B_1(\omega,m)&=&B_1\left(1-\omega,-m-\frac{2}{d}\right)\ ,
\nonumber\\
C_1(\omega,m)&=&C_1\left(1-\omega,-m-\frac{2}{d}\right)\ .
\label{duality}
\eea
To trace the origin of this invariance, we note that it
is present also in the Hessian.
More precisely, it is an invariance of the gauge-fixed spin-two, spin-one
and ghost kinetic operators.
It is not an invariance of the two-by-two spin-zero gauge-fixed Hessian matrix, but it is an invariance of its determinant.
The redefinition (\ref{betaredef}) is essential in order to
have the duality manifest.
The invariance of the Hessian implies that not only the one-loop divergences,
but the whole one-loop effective action
is invariant under (\ref{duality}).

There is a completely general
proof of the invariance of the measure under these transformations.
Assume that the quantum field is related to the metric by
$\gamma_{\mu\nu}=g_{\mu\nu}(\det g)^{w/2}$.
The relation between the variations of the
quantum field and the metric is then
\be
\delta \gamma_{\mu\nu}=
\delta\left(g_{\mu\nu}(\det g)^{w/2}\right)
=(\det g)^{w/2}
M_{\mu\nu}^{\alpha\beta}\delta g_{\alpha\beta}\ .
\ee
where
\be
M_{\mu\nu}^{\alpha\beta}=\left(\delta_{(\mu}^\alpha\delta_{\nu)}^\beta
+\frac{w}{2} g_{\mu\nu} g^{\alpha\beta}\right).
\ee
This can be inverted to give
\be
\delta g_{\alpha\beta}=
(\det g)^{-w/2}
\left(\delta^{\rho}_{(\alpha}\delta^{\sigma}_{\beta)}
-\frac{w/2}{1+dw/2}g^{\rho\sigma}g_{\alpha\beta}\right)
\delta\left(g_{\rho\sigma}(\det g)^{w/2}\right)\ .
\label{tr1}
\ee
Likewise, if the quantum field is related to the metric by
$\gamma^{\mu\nu}=g^{\mu\nu}(\det g)^{w'/2}$, we have
\be
\delta\gamma^{\mu\nu}=
\delta\left(g^{\mu\nu}(\det g)^{w'/2}\right)
=(\det g)^{w'/2}
\left(-g^{(\mu|\alpha}g^{|\nu)\beta}
+\frac{w'}{2} g^{\mu\nu}g^{\alpha\beta}\right)
\delta g_{\alpha\beta}\ .
\label{tr2}
\ee
Substituting (\ref{tr1}) into (\ref{tr2}) one finds
the Jacobian matrix for the change of variables:
\be
\frac{\delta\left(g^{\mu\nu}(\det g)^{w'/2}\right)}
{\delta\left(g_{\rho\sigma}(\det g)^{w/2}\right)}
=
(\det g)^{\frac{w'-w}{2}}
\left(-g^{(\mu|\rho}g^{|\nu)\sigma}
+\frac{w+w'}{2+dw}g^{\mu\nu}g^{\rho\sigma}\right).
\ee
The Jacobian determinant is
\be
\det\left(\frac{\delta\left(g^{\mu\nu}(\det g)^{w'/2}\right)}
{\delta\left(g_{\rho\sigma}(\det g)^{w/2}\right)}\right)=
\EuScript{N}
(\det g)^{\frac{w'-w}{2}\frac{d(d+1)}{2}-(d+1)}\ ,
\ee
where $\EuScript{N}$ is a numerical coefficient. The two measures are equivalent if this Jacobian determinant
is a purely numerical factor.
This happens if 
\be
\frac{w'}{2}=\frac{w}{2}+\frac{2}{d}\ .
\label{duality2}
\ee
Using (\ref{relation}), this is equivalent to $m'=-m-\frac{2}{d}$.
Thus, we have a general formal proof that if the action,
written in terms of $g_{\mu\nu}$, is kept fixed,
then the quantum theories defined in terms of the
densities 
$\gamma_{\mu\nu}=g_{\mu\nu}\sqrt{\det(g_{\mu\nu})}^w$
and
$\gamma^{\mu\nu}=g^{\mu\nu}\sqrt{\det(g_{\mu\nu})}^{w'}$,
with $w$ and $w'$ related as in (\ref{duality2}),
are equivalent.
The calculations we have reported in the previous sections
are a detailed confirmation of this statement.

We note that the relation (\ref{duality2}) has the following meaning:
if we give the metric any dimension $D$, then the variables
$\gamma_{\mu\nu}$ and $\gamma^{\mu\nu}$ have dimensions
$D(1+dw/2)$ and $D(-1+dw'/2)$ respectively,
and these dimensions agree if and only if (\ref{duality2}) holds.

In \cite{Fujikawa:1983im,Fujikawa:1984qk}, BRST invariance
was used as a criterion to fix $m$.
This leads to the two choices
\begin{equation}
\gamma_{\mu\nu}=g_{\mu\nu}(\det g)^{\frac{d-4}{4d}}\ ;
\qquad
\gamma^{\mu\nu}=g^{\mu\nu}(\det g)^{\frac{d+4}{4d}}\,.
\end{equation}
that correspond to
\be
m=\frac{4-d}{d^2}\ ;\qquad
m'= -\frac{d+4}{d^2}\ .
\ee
It is easy to check they are related through the duality transformation.

\section{Discussion}

We have investigated the dependence of the one-loop divergences
in Einstein theory on the choice of gauge and parametrization.
To avoid misunderstandings, we reiterate that whereas gauge dependence
is certainly unphysical, and must therefore drop out in any
observable, our treatment of different parametrizations
amounts really to different choices of functional measure.
Our analysis does not prove that observables can depend upon this
choice, but it shows at least that the divergent part of the effective action does. We have found that measures come in dual pairs
that lead to equivalent results. The implications of this result
for quantum gravity will have to be investigated more thoroughly.

One of the motivations for this work was to minimize
the gauge- and parametrization-dependence of the gravitational
beta functions and their fixed point.
In the context of asymptotic safety,
the coefficients $A_1$ and $B_1$ determine the beta functions
of $\Lambda$ and $G$.
In particular, Newton's constant has a fixed point at
$\tilde G_*=-(d-2)/B_1$.
For the exponential parametrization, $\tilde G_*$ is positive in any gauge as long as $a<3$.
In any parametrization, it vanishes along the line $b=3$, where $B_1$ diverges,
and there is a region near the line $b=3$ where it changes sign,
but we have seen that this pathological behavior 
can be attributed to a failure of the gauge-fixing.
The generally weak gauge-dependence is encouraging.

On the other hand, in any fixed gauge, $B_1$ has a strong
dependence on the parameters $m$ and $\omega$, such that $\tilde G_*$
becomes negative when $m$ and $\omega$ become simultaneously sufficiently large or small (upper right and lower left corners in Fig.~\ref{fig.fdd} and \ref{fig.phys}). For example, in $d=4$, this happens for
$\omega<1/2$ and $m<-\frac{7-4\omega}{8(1-2\omega)}$
or $\omega>1/2$ and $m>-\frac{7-4\omega}{8(1-2\omega)}$.
The origin of this parametrization-dependence is clear:
we did not take into account the Jacobians due to the changes of variables.
Then, different parametrizations really correspond to different definitions
of the functional integral, and hence in principle to different
quantum theories, so the observed parametrization-dependence
is not only acceptable but even expected.

Still, lacking strong arguments in favor of one specific choice of
measure, one may want to minimize the dependence of the results
on this choice.
We have seen that the choice $\omega=1/2$ 
(exponential parametrization)
and $b\to\pm\infty$ (unimodular gauge)
automatically eliminates also all dependence
on $m$, $a$ and on the cosmological constant.
Each of these quantities has a different physical meaning,
but each in its different way is a sources of uncertainties 
\footnote{In particular,
it has been known since long \cite{Reuter:2001ag}
that the appearance of the cosmological constant 
in the beta functions gives rise to singularities 
that prevent the smooth continuation of the
RG trajectories in the infrared.
The way this is solved in the exponential parametrization
has been discussed in \cite{pv1}.}.
A quantization scheme that eliminates these dependences
is therefore quite attractive
\footnote{The resulting effective action is in a sense 
closer to the on-shell effective action
\cite{bene,falls}}.

If one wants to minimize the dependence on the measure, 
a special choice clearly
stands out: it is the point $\omega=1/2$, $m=-1/d$,
or in other words the unimodular theory in exponential parametrization.
This is the unique point that is invariant under duality transformations,
and the unique stationary point for the coefficients $B_1$ and $C_1$.
We recall that in the case $m=-1/d$ the correspondence between
$g_{\mu\nu}$ and $\gamma_{\mu\nu}$ is not invertible,
so that the calculation presented here cannot be strictly
applied in that case.
This case has been considered from different viewpoints in
\cite{eichhorn,Saltas:2014cta,Alvarez:2015sba,Bufalo:2015wda,Benedetti:2015zsw}. Our results are a strong motivation to further investigate the quantum properties of this theory.

\section*{Acknowledgment}
The work of N.O. was supported in part by the JSPS Grants-in-Aid for Scientific Research (C) No. 24540290 and 16K05331. A.D.P. is grateful to Henrique Gomes, Flavio Mercati, Lee Smolin and specially Dario Benedetti for discussions. A.D.P. thanks CNPq, SISSA and DAAD for financial support and SISSA, Perimeter Institute and Albert Einstein Institute for hospitality.  This research was supported in part by Perimeter Institute for Theoretical Physics. Research at Perimeter Institute is supported by the Government of Canada through Industry Canada and by the Province of Ontario through the Ministry of Research and Innovation.

\appendix

\section{Some technical details}

Here we list the first three heat kernel coefficients for the
operator $\Delta=-\bnabla^2$ on a non-compact maximally symmetric space,
acting on spin-zero, spin-one and spin-two fields:
\bea
b_0(\Delta_{0})&=&1 \ ,
\nonumber\\
b_2(\Delta_{0})&=&\frac{1}{6}\bR \ ,
\nonumber\\
b_4(\Delta_{0})&=&\frac{6-7d+5d^2}{360d(d-1)}\bR^2 \ ,
\nonumber\\
b_0(\Delta_{1})&=& d-1 \ ,
\nonumber\\
b_2(\Delta_{1})&=&\left(\frac{d-1}{6}-\frac{1}{d}\right)\bR \ ,
\nonumber\\
b_4(\Delta_{1})&=&\frac{180-186d-47d^2-12d^3+5d^4}{360d^2(d-1)}\bR^2 \ ,
\nonumber\\
b_0(\Delta_{2})&=&\frac{(d+1)(d-2)}{2} \ ,
\nonumber\\
b_2(\Delta_{2})&=&\frac{(d+1)(d+2)(d-5)}{12(d-1)}\bR \ ,
\nonumber\\
b_4(\Delta_{2})&=&\frac{(d+1)(-228-392d-83d^2-22d^3+5d^4)}{720d(d-1)^2}\bR^2 \ ,
\eea
In the compact case (a sphere) there are some discrete modes that have
to be removed from the spectrum and change the coefficient of $\bR^2$.
In this paper we restrict our attention to the non-compact case,
where the spectrum is continuous.
Then, the modes to be removed are of measure zero and have no effect.

Finally we list the values of the coefficients that enter in (\ref{euterpe})
and its lower-spin analogues:
\bea
Q_{d/2}&=&\frac{2k^d}{d\Gamma(d/2)} \ ,
\nonumber\\
Q_{d/2-1}&=&\frac{k^{d-2}}{\Gamma(d/2)} \ ,
\nonumber\\
Q_{d/2-2}&=&\frac{(d-2)k^d}{2\Gamma(d/2)}\ .
\eea

\section{Comparison with the literature}

The formula (\ref{C1lin}) can be compared with an old calculation
of Kallosh et al. \cite{kallosh}, giving
\be
C_1=\frac{1}{8\pi^2}\left(\frac{3}{2}a+\frac{1}{4}b\right)\ ,
\nonumber
\ee
where $a$ and $b$ (not to be confused with our gauge parameters) 
are given in their equations (2.10) and (2.11).
Taking into account that their gauge parameters
$a_K$ and $b_K$ are related to ours by
$a_K=-1/a$, $b_K=-(1+b)/d$, this translates to
\bea
C_1&=&\frac{1}{17280\pi^2(b-3)^4}
\Big[135a^2\left(3b^4-36b^3+162b^2-324b+259\right)
\nonumber\\
&&
\qquad\qquad\qquad\qquad
-180a\left(3b^4-36b^3+176b^2-360b+297\right)
\nonumber\\
&&
\qquad\qquad\qquad\qquad
+216\left(7b^4-59b^3+223b^2-381b+252\right)\Big].
\eea
The difference with our result is $53/4320\pi^2$,
which corresponds to the term in the $B_4$ coefficient
proportional to the Euler invariant.
Up to this irrelevant total derivative term,
there is therefore complete agreement.

The second paper in \cite{Kalmykov:1995fd} contains a more general calculation,
which corresponds in our notation to the cases $\omega=0$ or
$\omega=1$, with arbitrary $m$.
In the case $\omega=0$ our result for $C_1$ is
\bea
C_1&=&
\frac{1}{17280\pi^2(b-3)^4(1+ 4 m)^2}\Big\{
324(221+3958 m+24236m^2+57600m^3+43200m^4)
\nonumber\\&&\qquad\qquad\qquad\qquad
-216\, b\, (1 + 4 m) (487 + 6388 m + 23400 m^2 + 21600 m^3)
\nonumber\\&&\qquad\qquad\qquad\qquad
+432\, b^2 (1 + 4 m) (138 + 1697 m + 5910 m^2 + 5400 m^3)
\nonumber\\&&\qquad\qquad\qquad\qquad
-24\, b^3 (1 + 4 m)^2 (637 + 5040 m + 5400 m^2)
\nonumber\\&&\qquad\qquad\qquad\qquad
+4\, b^4 (1 + 4 m)^2 (431 + 3510 m + 4860 m^2)
\nonumber\\&&\qquad\qquad\qquad
+a \big[ - 1620 (33 + 404 m + 1696 m^2 + 2304 m^3)
\nonumber\\&&\qquad\qquad\qquad\qquad
+ 12960\, b\, (1 + 4 m) (5 + 42 m + 92 m^2)
\nonumber\\&&\qquad\qquad\qquad\qquad
- 1440\, b^2(1 + 4 m) (22 + 185 m + 426 m^2 + 144 m^3)
\nonumber\\&&\qquad\qquad\qquad\qquad
-540\,  b^3 (b-12)(1 + 4 m)^3  
\big]
\nonumber\\&&\qquad\qquad\qquad
+a^2\big[135 (259 + 4144 m + 25120 m^2 + 68352 m^3 + 71424 m^4)
\nonumber\\&&\qquad\qquad\qquad\quad\quad
+405\, b\,(b-6) (18 - 6 b + b^2) (1 + 4 m)^4
\big]
\Big\}\ .
\label{C1gen0}
\eea
This should be compared to the quantity
\be
\frac{1}{16\pi^2}\left(\frac{c_1}{4}+c_4\right)
\label{pronin}
\ee
where $c_1$ and $c_4$ are given in their equations (20)-(21),
with $f_1=\frac{1}{1+4r}$, $f_2=-\frac{r}{1+4r}$,
$f_3=f_4=1$, $f_5=0$, $f_6=\frac{6r^2+4r+1}{(1+4r)^2}$.
Their gauge parameters are related to ours by
$\alpha=a-3$, $\gamma=\frac{2}{3-b}$
and $r$ is equal to $w/2$, which is related to $m$ as in the first equation in (\ref{relation}).
When this is done, the result is again found to differ from
(\ref{C1gen0}) by the Euler term $53/4320\pi^2$.

In the case $\omega=1$ our result for $C_1$ is
\bea
C_1&=&
\frac{1}{17280\pi^2(b-3)^4(1+ 4 m)^2}\Big\{
324 (-199 - 1322 m + 2636 m^2 + 28800 m^3 + 43200 m^4)
\nonumber\\&&\qquad\qquad\qquad\qquad
-216\, b\, (1 + 4 m) (-443 - 812 m + 9000 m^2 + 21600 m^3)
\nonumber\\&&\qquad\qquad\qquad\qquad
+ 432\, b^2 (1 + 4 m) (-92 - 163 m + 2190 m^2 + 5400 m^3)
\nonumber\\&&\qquad\qquad\qquad\qquad
- 24\, b^3 (1 + 4 m)^2 (-533 + 360 m + 5400 m^2) 
\nonumber\\&&\qquad\qquad\qquad\qquad
+4\, b^4 (1 + 4 m)^2 (-109 + 1350 m + 4860 m^2) 
\nonumber\\&&\qquad\qquad\qquad
+a\big[ 1620 (33 + 436 m + 1760 m^2 + 2304 m^3)
\nonumber\\&&\qquad\qquad\qquad\qquad
- 12960\, b\,(1 + 4 m) (7 + 50 m + 92 m^2)
\nonumber\\&&\qquad\qquad\qquad\qquad
+ 1440\, b^2 (1 + 4 m) (18 + 133 m + 210 m^2 - 144 m^3)
\nonumber\\&&\qquad\qquad\qquad\qquad
+540\, b^3 (b-12) (1 + 4 m)^3
\big]
\nonumber\\&&\qquad\qquad\qquad
+a^2\big[135 (387 + 5424 m + 29728 m^2 + 74496 m^3 + 71424 m^4)
\nonumber\\&&\qquad\qquad\qquad\qquad
+405\, b\, (b-6)  (18 - 6 b + b^2) (1 + 4 m)^4
\big]
\Big\}\ .
\eea
This time we compare again with (\ref{pronin}) but
with $f_1=\frac{1}{1-4p}$, $f_2=\frac{p}{1-4p}$,
$-f_3=f_5=1$, $f_4=0$, $f_6=\frac{6p^2-4p+1}{(1-4p)^2}$.
and $p=w/2$, related to $m$ as in the second equation in (\ref{relation}).
The results agree again modulo $53/4320\pi^2$.

\end{document}